 \documentclass[twocolumn]{aastex6}

\usepackage{color}

\newcommand{\bwt}{\begin{widetext}}
\newcommand{\ewt}{\end{widetext}}
\newcommand{\beq}{\begin{equation}}
\newcommand{\eeq}{\end{equation}}
\newcommand{\bea}{\begin{eqnarray}}
\newcommand{\eea}{\end{eqnarray}}

\begin{document}
                    
\title{Cooling of Small and Massive Hyperonic Stars}

\author{Rodrigo Negreiros}
\affil{Instituto de F\'isica, Universidade Federal Fluminense, Av. Gal. Milton Tavares S/N, Niter\'oi, Brazil}

\author{Laura Tolos}
\affil{Institut f\"ur Theoretische Physik, Goethe Universit\"at Frankfurt, Max von Laue Strasse 1, 60438 Frankfurt, Germany\\
Frankfurt Institute for Advanced Studies,  Goethe Universit\"at Frankfurt, \\ Ruth-Moufang-Str. 1, 60438 Frankfurt am Main, Germany \\
Institute of Space Sciences (ICE, CSIC), Campus UAB, Carrer de Can Magrans, 08193, Barcelona, Spain \\
Institut d'Estudis Espacials de Catalunya (IEEC), 08034 Barcelona, Spain
}

\author{Mario Centelles}
\affil{Departament de F\'{\i}sica Qu\`antica i Astrof\'{\i}sica and Institut de Ci\`encies del Cosmos
(ICCUB), Facultat de F\'{\i}sica, Universitat de Barcelona, Mart\'{\i} i Franqu\`es 1, 08028
Barcelona, Spain}

\author{Angels Ramos}
\affil{Departament de F\'{\i}sica Qu\`antica i Astrof\'{\i}sica and Institut de Ci\`encies del Cosmos
(ICCUB), Facultat de F\'{\i}sica, Universitat de Barcelona, Mart\'{\i} i Franqu\`es 1, 08028
Barcelona, Spain}

\author{Veronica Dexheimer}
\affiliation{Department of Physics, Kent State University, Kent OH 44242 USA}

\begin{abstract}
We perform cooling simulations for isolated neutron stars using recently developed equations of state for their core. The equations of state are obtained from new parametrizations of the FSU2 relativistic mean-field functional that reproduce the properties of nuclear matter and finite nuclei, while fulfilling the restrictions on high-density matter deduced from heavy-ion collisions, measurements of massive 2$M_{\odot}$ neutron stars, and neutron star radii below 13 km. We find that two of the models studied, FSU2R (with nucleons) and in particular FSU2H (with nucleons and hyperons), show very good agreement with cooling observations, even without including extensive nucleon pairing. 
This suggests that the cooling observations are more compatible with an equation of state that produces a soft nuclear symmetry energy and, hence, generates small neutron star radii.  However, both models favor large stellar masses, above $1.8 M_{\odot}$, to explain the colder isolated neutron stars that have been observed, even if nucleon pairing is present.
\end{abstract}


\vspace*{5mm}
\section{Introduction}
\label{sec:intro}

The study of the thermal evolution of neutron stars has been a prominent tool for probing the equation of state (EoS) and composition of these objects \citep{TSURUTA1965,Maxwell1979,Page2006,Weber2007,Negreiros2010}. The reason behind this is that the thermal properties that govern the cooling down of neutron stars, particularly neutrino emission processes,  depend strongly on the particle composition and, thus, on the EoS of dense matter \citep{Page2006,Page2009}. Furthermore, recent observations have produced a wealth of neutron star data that can be used to constrain the properties of the underlying microscopic models used to describe these objects. 

Several works have addressed the thermal evolution of neutron stars, comparing it with the observed data, under the light of different phenomena, such as deconfinement to quark matter \citep{Horvath1991,Blaschke2000a,Shovkovy2002,Grigorian2005,Alford2005a}, stellar rotation \citep{Negreiros2012,Negreiros2013,Negreiros2017}, superfluidity \citep{Levenfish1994,Schaab1997,Alford2005,Page2009,Fortin:2017rxq}, magnetic fields \citep{Aguilera2008,Pons2009,Niebergal2010a,Negreiros2017b}, among others \citep{Weber2005,Alford2005,Gusakov2005,Negreiros2010}. 
A main point of contention in the thermal evolution studies is whether or not fast cooling processes take place, because if the star cools down too fast, it will yield to disagreement with most observed data \citep{Page2004}. The most prominent fast cooling process is the direct URCA (DU) process, i.e., the beta decay of a neutron and the electron capture by a proton.  These processes are present if the proton fraction is high enough as to allow for momentum conservation. This leads to a direct connection between the thermal behavior and the symmetry energy of nuclear matter, as the latter is directly related to the proton fraction \citep{Boguta:1981mw,Horowitz:2002mb,Steiner:2004fi,Than:2009ct,Beloin:2016zop}.

In this work, we investigate the thermal evolution of neutron stars whose composition is described by the microscopic models developed in \cite{Tolos:2016hhl,Tolos:2017lgv}, which produce EoS's for the nucleonic and hyperonic inner core of neutron stars that reconcile the $2 M_{\odot}$ mass observations  
\citep{Demorest:2010bx,Antoniadis:2013pzd} with the recent analyses of stellar radii below 13~km \citep{Guillot:2013wu,Guillot:2014lla,Guver:2013xa,Heinke:2014xaa,Lattimer:2014sga,Lattimer:2013hma,Nattila:2015jra,Ozel:2015fia,Ozel:2016oaf,Lattimer:2015nhk},
and therefore satisfy the upper limit of $\sim$13.4--13.8~km for the radius of a $1.4 M_\odot$ neutron star that
has been recently deduced \citep{Fattoyev:2017jql,Annala:2017llu,Krastev:2018nwr,Most:2018hfd}
from the gravitational-wave signal of a neutron-star binary merger detected by the LIGO and Virgo collaborations \citep{TheLIGOScientific:2017qsa}.
Moreover, the properties of nuclear matter and of finite nuclei  \citep{Tsang:2012se,Chen:2014sca} are reproduced together with the constraints on the EoS extracted from nuclear collective flow \citep{Danielewicz:2002pu} and kaon production \citep{Fuchs:2000kp,Lynch:2009vc} in heavy-ion collisions (HICs).  The study is performed within the relativistic mean-field (RMF) theory for describing both the nucleon and hyperon interactions and the EoS of the neutron star core. 

Two models have been formulated, denoted by FSU2R (with nucleons) and FSU2H (with nucleons and hyperons), based on the nucleonic FSU2 model of \cite{Chen:2014sca}.  For the FSU2H model, the couplings of the mesons to the hyperons are fixed to reproduce the available hypernuclear structure data. 
The impact of the experimental uncertainties of the hypernuclear data on the stellar properties was analyzed in \cite{Tolos:2017lgv}. The main differences between the two models were found in the onset of appearance of each hyperon. However, the values of the neutron star maximum masses showed only a moderate dispersion of about~0.1$M_{\odot}$.  Note that a broader dispersion of values may be expected from the lack of knowledge of the hyperon-nuclear interaction at the high densities present in the center of 2$M_{\odot}$ stars. Hopefully, advances in the theoretical description of hyperon-nucleon interactions in dense matter from chiral effective forces \citep{Haidenbauer:2016vfq} and possible constraints from future HIC experiments \citep{Ohnishi:2016elb} will help narrowing down these uncertainties.

In the present study, we focus on how the hyperons as well as the symmetry energy of the microscopic model influence the cooling history of neutron stars 
considering different nucleonic pairing scenarios. However, we do not consider hyperonic pairing,  stellar rotation nor  magnetic fields \citep{Negreiros2012,Negreiros2013,Negreiros2017}, which are left for future study.
We note that after hyperons were first proposed as one of the possible components of neutron stars \citep{Glendenning:1982nc}, their possible influence on the cooling of neutron stars was  investigated in \cite{Prakash1992}, and included in most thermal evolution studies thereafter. Most recently, \cite{2018MNRAS.475.4347R} revisited the topic by  analyzing the cooling of massive stars described by different relativistic density functional models including nucleonic and hyperonic pairing. We will show that the microscopic models proposed here, given their underlying properties (especially the density slope of the symmetry energy), provide a very satisfactory agreement with observed cooling data. 
Particularly interesting is the fact that even without resorting to proton pairing deep in the inner core, the results agree with most cooling observations for a large range of neutron stars.


\vspace*{5mm}
\section{The Models}
\subsection{Underlying Lagrangian}
\label{sec:formalism}

Our models are based on two new parametrizations of the FSU2 RMF model  
\citep{Chen:2014sca}. The Lagrangian density of the theory reads 
\citep{Serot:1984ey,Serot:1997xg,Glendenning:2000,Chen:2014sca}
\beq
{\cal L}= \sum_{b}{\cal L}_{b} + {\cal L}_{m}+ \sum_{l=e, \mu}{\cal L}_{l} \ ,
\label{lan}
\eeq
with the baryon ($b$), meson ($m$=$\sigma$, $\omega$, $\rho$ and $\phi$), and lepton ($l$=$e$, $\mu$) Lagrangians given by
\bea
{\cal L}_{b}&=&\bar{\Psi}_{b}(i\gamma_{\mu}\partial^{\mu}-m_{b}  +  g_{\sigma b}\sigma-g_{\omega b}\gamma_{\mu}\omega^{\mu}-g_{\phi b}\gamma_{\mu}\phi^{\mu}\nonumber \\
&-&g_{\rho b}\gamma_{\mu}\vec{I}_b  \vec{\rho \, }^{\mu} 
)\Psi_{b} , \nonumber \\[2mm] 
{\cal L}_{m}&=&\frac{1}{2}\partial_{\mu}\sigma \partial^{\mu}\sigma
-\frac{1}{2}m^{2}_{\sigma}\sigma^{2} - \frac{\kappa}{3!} (g_{\sigma N}\sigma)^3 - \frac{\lambda}{4!} (g_{\sigma N}\sigma)^4 \nonumber \\
&-& \frac{1}{4}\Omega^{\mu \nu} \Omega_{\mu \nu} +\frac{1}{2}m^{2}_{\omega}\omega_{\mu}\omega^{\mu}  + \frac{\zeta}{4!}   (g_{\omega N}\omega_{\mu} \omega^{\mu})^4 \nonumber \\
&-&\frac{1}{4}  \vec{R}^{\mu \nu}\vec{R}_{\mu \nu}+\frac{1}{2}m^{2}_{\rho}\vec{\rho}_{\mu}\vec{\rho \, }^{\mu} + \Lambda_{\omega} g_{\rho N}^2 \vec{\rho}_{\mu}\vec{\rho \,}^{\mu} g_{\omega N}^2 \omega_{\mu} \omega^{\mu} \nonumber \\
&-&\frac{1}{4}  P^{\mu \nu}P_{\mu \nu}+\frac{1}{2}m^{2}_{\phi}\phi_{\mu}\phi^{\mu} \ ,\nonumber \\[2mm] 
{\cal L}_{l}&=& \bar{\psi}_{l}\left(i\gamma_{\mu}\partial^{\mu}
-m_{l}\right )\psi_{l} \ ,
\label{lagran}
\eea
where $\Psi_{b}$ and $\psi_{l} $ are the baryon and lepton Dirac fields, respectively.  The field strength tensors are 
$\Omega_{\mu \nu}=\partial_{\mu}\omega_{\nu}-\partial_{\nu}\omega_{\mu}$, $\vec{R}_{\mu \nu}=\partial_{\mu}\vec{\rho}_{\nu}-\partial_{\nu}\vec{\rho}_{\mu} $, and $P_{\mu \nu}=\partial_{\mu}\phi_{\nu}-\partial_{\nu}\phi_{\mu}$. 
The strong interaction couplings of a meson to a certain baryon are denoted by $g$ and the baryon, meson and lepton masses by $m$.  The vector $\vec{I}_b$ stands for the isospin operator.

The Lagrangian density (\ref{lagran}) incorporates scalar and vector meson 
self-interactions, as well as a mixed quartic vector meson interaction. The nonlinear 
meson interactions are important to describe nuclear matter 
and finite nuclei. The scalar self-interactions with coupling constants $\kappa$ and $\lambda$ \citep{Boguta:1977xi} are responsible for 
softening the EoS of symmetric nuclear matter around saturation density and allow 
one to obtain a realistic value for the compression modulus of nuclear matter 
\citep{Boguta:1977xi,Boguta:1981px}. The quartic isoscalar-vector self-interaction 
(with coupling $\zeta$) softens the EoS at high densities \citep{Mueller:1996pm}, 
while the mixed quartic isovector-vector interaction (with coupling 
$\Lambda_{\omega}$) is introduced \citep{Horowitz:2000xj,Horowitz:2001ya} to modify 
the density dependence of the  nuclear symmetry energy.
From the Lagrangian density (\ref{lagran}), one derives in the standard way the equations
of motion for the baryon and meson fields, which are solved in the mean-field approximation
\citep{Serot:1984ey}.

To compute the structure of neutron stars, one needs the EoS of matter over a wide range of densities. 
The core of a neutron star harbours chemically-equilibrated ($\beta$-stable), globally neutral, charged matter. 
We compute the EoS for the core of the star using the Lagrangian of Eqs.~(\ref{lan})--(\ref{lagran}).
As in \cite{Tolos:2016hhl,Tolos:2017lgv}, for the crust of the star we employ the crustal EoS of \cite{Sharma:2015bna}, which has been obtained from microscopic calculations. 
Once the EoS is known, the solution of the Tolman-Oppenheimer-Volkoff (TOV) equations \citep{Oppenheimer:1939ne} provides the mass $M$ and radius $R$ of the neutron star.

\subsection{Parametrizations}
\label{sec:calibration}

\begin{table}[t]
\begin{center}
\begin{tabular}{|c|c|c|c|}
\hline 
Models &  FSU2  & FSU2R  & FSU2H \\ 
\hline
$m_{\sigma}$ (MeV)  &   497.479& 497.479  &  497.479 \\

$m_{\omega}$ (MeV) &  782.500& 782.500  &  782.500 \\

$m_{\rho}$ (MeV) &   763.000& 763.000 &  763.000 \\

$g_{\sigma N}^2$ &   108.0943& 107.5751 &  102.7200 \\

$g_{\omega N}^2$  &  183.7893& 182.3949  &  169.5315\\

$g_{\rho N}^2$ &  80.4656& 206.4260  &  197.2692 \\

$\kappa$  &  3.0029& 3.0911  &  4.0014 \\

$\lambda$ &  $-$0.000533& $-$0.001680  &  $-$0.013298\\

$\zeta$ &  0.0256& 0.024  &  0.008 \\

$\Lambda_{\omega}$  &  0.000823& 0.045  &  0.045 \\
\hline
$n_0$ $({\rm fm}^{-3})$  & 0.1505  & 0.1505   & 0.1505 \\

$E/A$ $({\rm MeV})$  & $-$16.28 & $-$16.28 & $-$16.28 \\

$K$ $({\rm MeV})$  &  238.0 &  238.0 &  238.0  \\


$E_{\rm sym}(n_0)$ (MeV) &   37.6&  30.7  &  30.5 \\

$L$ (MeV)   &  112.8& 46.9  &  44.5 \\


\hline
 \end{tabular}
\end{center}
\caption{Parameters of the models FSU2 \citep{Chen:2014sca}, and FSU2R and FSU2H  
\citep{Tolos:2016hhl,Tolos:2017lgv} [here we use the slightly updated version of the 
FSU2R and FSU2H parameters given in \citep{Tolos:2017lgv}]. 
The mass of the nucleon is $m=939$ MeV.
Also reported are the values in nuclear matter at the saturation density $n_0$
for the energy per particle ($E/A$), incompressibility ($K$), symmetry energy 
($E_{\rm sym}$), and slope parameter of the symmetry energy~($L$).}
\label{t-parameters}
\end{table}

\begin{figure}[t]
\centering
\includegraphics[width=0.9\columnwidth]{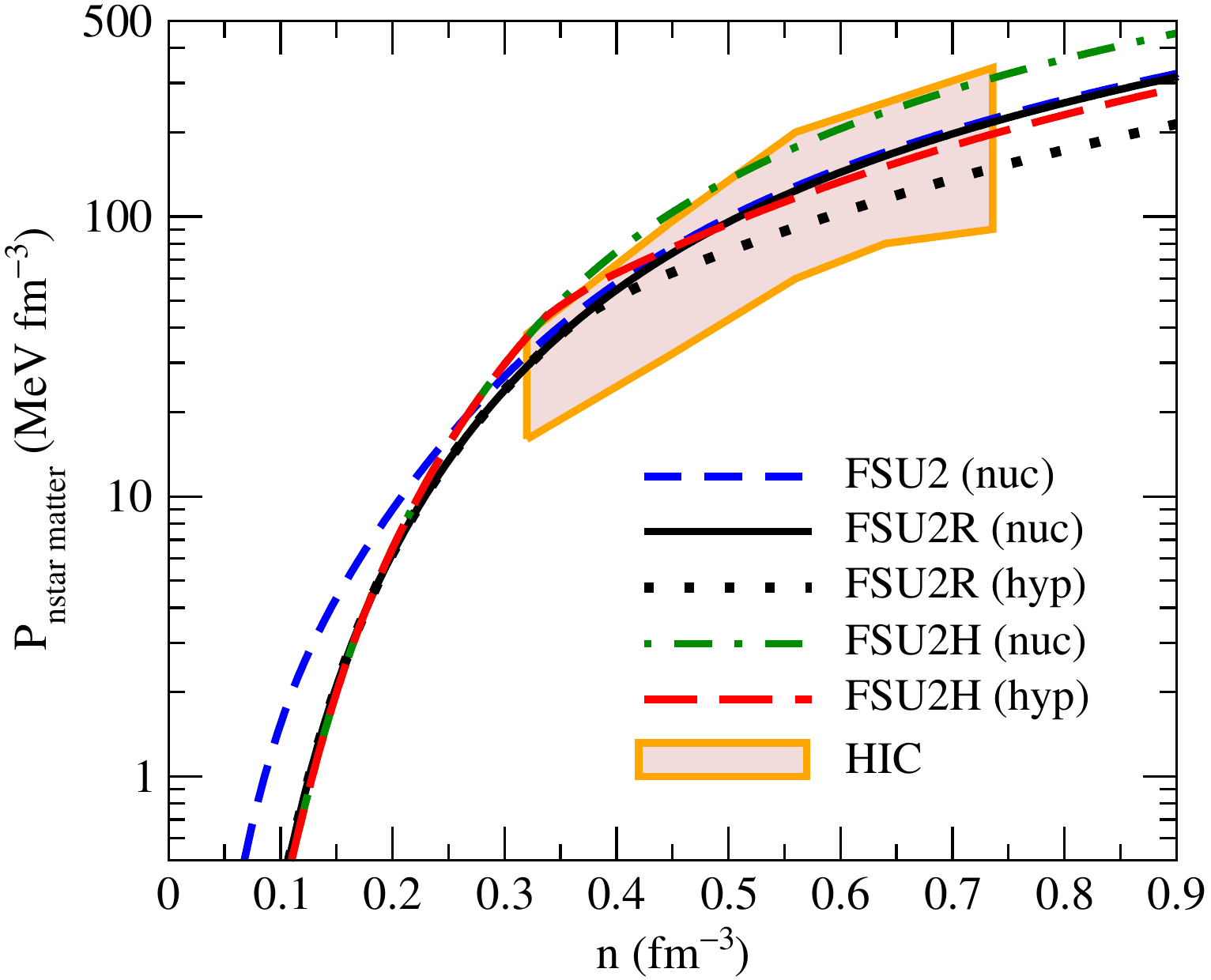}
\caption{Pressure of $\beta$-stable neutron star matter as a function of baryon 
number density for the models used in this work. 
The colored area depicts the region compatible with collective flow observables in 
energetic heavy-ion collisions \citep{Danielewicz:2002pu} (we note that although this constraint was 
deduced for pure neutron matter, it is useful here because at the implied 
densities the pressures of neutron matter and $\beta$-stable matter are very similar).}
\label{fig:eosbeta}
\end{figure}

\begin{figure}[t]
\centering
\includegraphics[width=0.9\columnwidth]{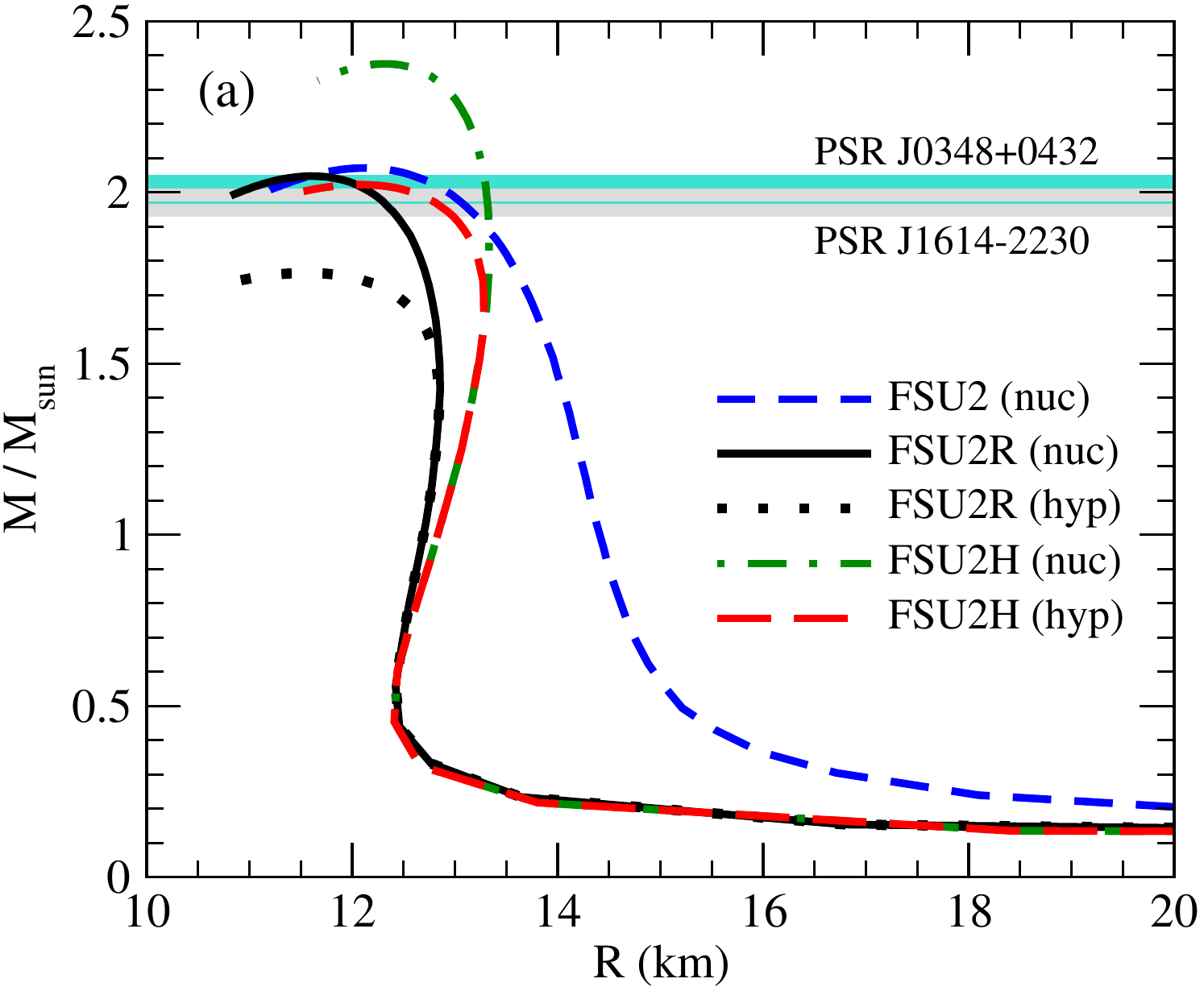}
\vspace*{0.mm}

\includegraphics[width=0.9\columnwidth]{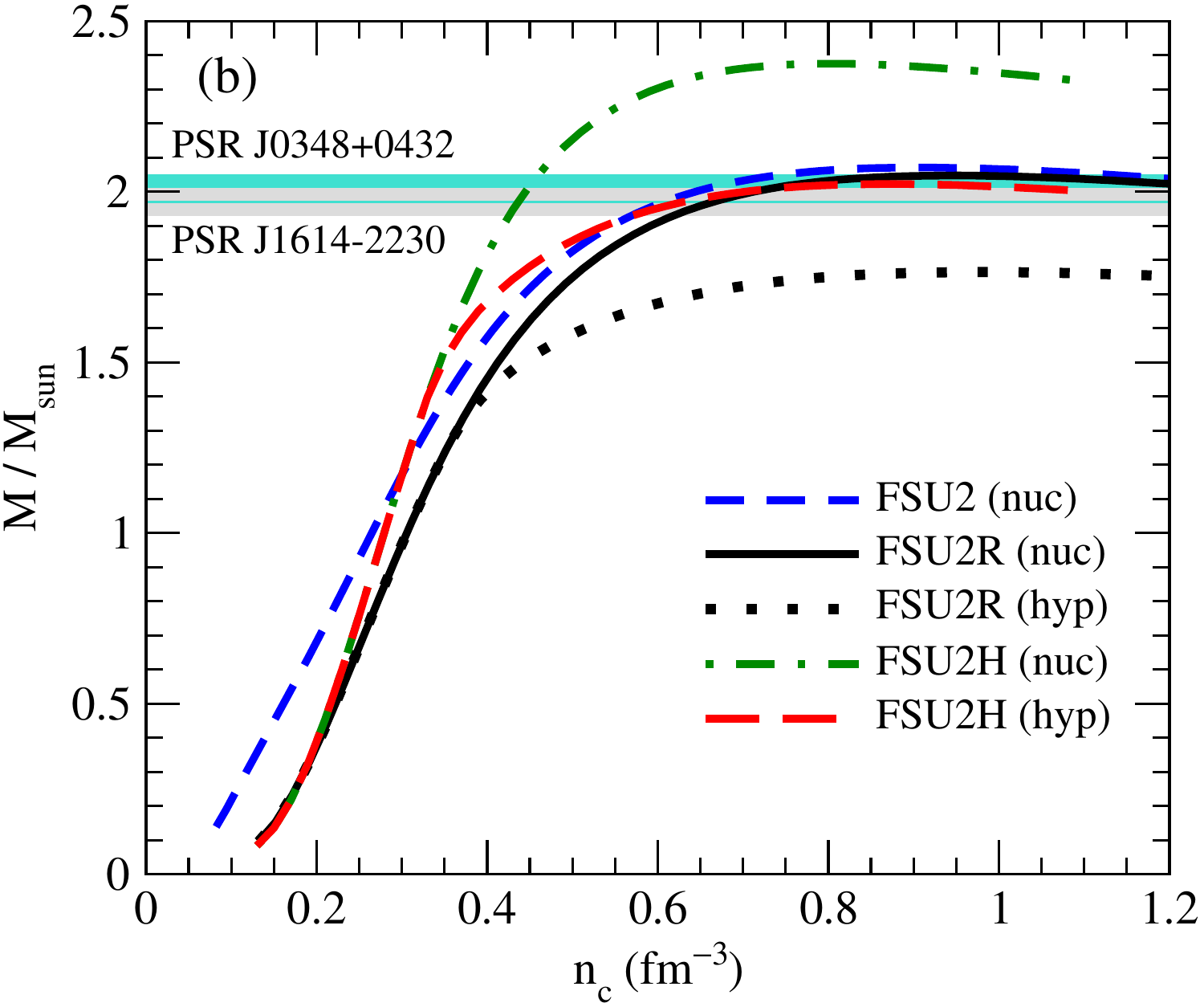}
\caption{{(a) Mass-radius relation for neutron stars predicted by the models used in this 
work. (b) Mass of neutron stars versus central density for the same models.}
The accurate mass measurements of $1.97 \pm 0.04 M_\odot$ in pulsar PSR 
J1614--2230 \citep{Demorest:2010bx} and $2.01 \pm 0.04 M_\odot$ in pulsar PSR 
J0348+0432 \citep{Antoniadis:2013pzd} are also shown.}
\label{fig:MRplot}
\end{figure}

We first consider the nucleonic RMF parametrization FSU2 of \cite{Chen:2014sca}. The parameters of the model were fitted 
by requiring an accurate description of the binding energies, charge radii and monopole response of atomic 
nuclei across the periodic table, and, at the same time, a limiting mass of $\simeq$\,$2 M_\odot$ in neutron 
stars. The resulting FSU2 parameter set \citep{Chen:2014sca}
(the parameters and saturation properties of the models can be found in Table~\ref{t-parameters}) 
provides a stiff enough EoS in the region of large baryon densities and, as a 
consequence, reproduces heavy neutron stars. Figure~\ref{fig:eosbeta} shows the 
pressure of $\beta$-stable neutron star matter as a function of the baryon density for 
the different models considered in the present work. It can be seen that the pressure 
from FSU2---and also the pressure from all the other models that we use---passes 
through the region constrained by the study of flow data in experiments of energetic 
HICs \citep{Danielewicz:2002pu}. The predicted mass-radius (M-R) relation of neutron stars 
is represented in {the upper panel of Fig.~\ref{fig:MRplot}. The lower panel of Fig.~\ref{fig:MRplot}
shows the mass of the star as a function of the central density.} FSU2 reaches a 
maximum mass of $2.07 M_\odot$ and, thus, accommodates the observed masses 
of $2 M_\odot$ in pulsars PSR J1614--2230 \citep{Demorest:2010bx} and PSR J0348+0432 
\citep{Antoniadis:2013pzd}. The radius of the star at maximum mass is of 12.1 km in 
FSU2 (see Table~\ref{tab:starprops}). For a neutron star with a mass of $1.4 M_\odot$, 
FSU2 predicts a stellar radius of 14.1~km. 

Recent progress in astrophysical estimates of neutron star radii---see for example 
\cite{Ozel:2016oaf} for a review---suggests that radii may be smaller than previously 
thought. Particularly, the extractions of radii from quiescent low-mass X-ray binaries 
(QLMXBs) and X-ray bursters are pointing toward values no larger than approximately 13~km 
\citep{Guillot:2013wu,Guillot:2014lla,Guver:2013xa,Heinke:2014xaa,Lattimer:2014sga,
Lattimer:2013hma,Nattila:2015jra,Ozel:2015fia,Ozel:2016oaf,Lattimer:2015nhk}.
The first observation of a binary neutron-star merger by the LIGO and Virgo
collaborations also appears to indicate that neutron stars cannot have excessively large 
radii \citep{TheLIGOScientific:2017qsa}.
{Indeed, very recently, by comparing the tidal deformabilities extracted from the 
gravitational-wave signal of this merger with the predictions using different types of  
EoS's, relatively small upper limits for the radius of a $1.4 M_\odot$ neutron star have 
been deduced such as of $R<13.76$~km in \citep{Fattoyev:2017jql}, of $R<13.6$~km in 
\citep{Annala:2017llu,Krastev:2018nwr}, and of $R<13.45$~km in \citep{Most:2018hfd}.}
It is worth mentioning that the advent of accurate data on neutron star radii should 
allow one to probe the EoS of neutron-rich matter in a complementary way to the heavy masses. This is due to 
the fact that, while the maximum stellar mass depends on the high-density 
sector of the EoS, the strongest impact on the radius of the star comes from the 
EoS in the low-to-medium density region of \mbox{1--2}~times the 
saturation density $n_0$ \citep{Lattimer:2006xb,Ozel:2016oaf}. Given that the nuclear 
symmetry energy $E_{\rm sym}(n)$ governs the departure of the energy of neutron matter 
from symmetric matter, it means that data on neutron star radii pin down the 
density dependence of the symmetry energy around $n_0$ and the slope parameter $L$, defined 
as $\displaystyle L= 3 n_0 \left.\frac{\partial E_{\rm sym}(n)}{\partial n} \right\vert_{n=n_0}$, 
which is intimately related with the isospin properties of atomic nuclei, although its value is still uncertain \citep{Li:2014oda}. 
Larger $L$ values (stiffer symmetry energy) favor larger radii in neutron stars, whereas smaller $L$ values
(softer symmetry energy) favor smaller radii. {Therefore, astrophysical evidence of
small neutron star radii is consistent with a nuclear symmetry energy that is not overly 
stiff around saturation density.}

\begin{table*}[t]
\begin{center}
\begin{tabular}{|c|c|c|c|c|c|c|c|}
\hline
Models & $M_{\rm max}/M_\odot$ & $R(M_{\rm max})$ (km) & $n_c(M_{\rm max})/n_0$  & 
$R(1.4M_\odot)$ (km) &  $n_Y/n_0$ \\ 
\hline
FSU2(nuc) \ \ & {2.071} & 12.1 & 5.9 & 14.1  & --- \\
FSU2R(nuc)    & {2.048} & 11.6 & 6.3  & 12.8  & --- \\
FSU2H(nuc)    & {2.376} & 12.3 &  5.4  & 13.2  & --- \\ 
\hline
FSU2R(hyp)    & {1.765} & 11.6 & 6.5  & 12.8 &  2.4\\
FSU2H(hyp)    & {2.023} & 12.1 &  5.8 & 13.2 & 2.2\\ 
\hline 
 \end{tabular}
\end{center}
\caption{Neutron star properties from the models used in this work. Results 
are shown for nucleonic (nuc) and hyperonic (hyp) stars. 
The quantity $n_c(M_{\rm max})/n_0$ is the central baryon number density at the star with maximum mass, $M_{\rm max}$,
normalized to the saturation density $n_0$, whereas $n_Y/n_0$ is the 
onset density of appearance of hyperons normalized to $n_0$.}
\label{tab:starprops}
\end{table*}

To account for the possible existence of massive stars with small radii in our 
theory, yet without compromising the agreement with constraints from the properties of 
atomic nuclei and from HICs, in \cite{Tolos:2016hhl,Tolos:2017lgv} we developed the 
FSU2R parametrization, which produces a soft symmetry energy and a soft pressure of 
neutron matter for densities $n \lesssim 2n_0$. This can be seen in 
Fig.~\ref{fig:eosbeta} by comparing the EoS's of FSU2R (solid black line) and FSU2 
(dashed blue line). For a given neutron star, in the FSU2R EoS up to densities
$\sim$ $2n_0$ there is less pressure to balance gravity, thereby leading to increased 
compactness of the star and smaller stellar radius. In the high-density sector of the 
EoS, the FSU2R and FSU2 EoS's are close to each other (cf.\ Fig.~\ref{fig:eosbeta}), 
and, thus, FSU2R also reproduces heavy neutron stars, as can be seen in Fig.~\ref{fig:MRplot}.
Smaller radii within the range of 11.5--13~km are obtained in FSU2R for neutron stars between 
maximum mass and $M= 1.4 M_\odot$ (see  Fig.~\ref{fig:MRplot} and Table~\ref{tab:starprops}), owing to a softer 
symmetry energy, while reproducing the properties of nuclear matter and 
nuclei. Our conclusions are in keeping with the results of recent studies with RMF 
models with a soft symmetry energy \citep{Chen:2015zpa,Chen:2014mza}. Indeed, FSU2R 
predicts $E_{\rm sym}(n_0)= 30.7$ MeV and $L=46.9$ MeV \citep{Tolos:2017lgv}, 
which are in good accord with the limits of recent determinations 
\citep{Lattimer:2012xj,Li:2013ola,Roca-Maza:2015eza,Hagen:2015yea,Birkhan:2016qkr,Oertel:2016bki}.

As the FSU2 and FSU2R parametrizations assume nucleonic (non-strange) stellar cores, 
we shall often use the notation FSU2(nuc) and FSU2R(nuc) to refer to these models. 
We also analyze in the present work the consequences of the appearance of hyperons 
inside neutron star cores. 
As described in \cite{Tolos:2016hhl,Tolos:2017lgv}, the couplings of the hyperons to the 
vector mesons are related to the nucleon couplings by assuming SU(3)-flavor symmetry, 
the vector dominance model and ideal mixing for the physical $\omega$ and $\phi$ mesons, 
as, e.g., employed in recent works 
\citep{Schaffner:1995th,Banik:2014qja,Miyatsu:2013hea,Weissenborn:2011ut, 
Colucci:2013pya}. The values of the hyperon couplings to the scalar $\sigma$ meson field 
are determined from the available experimental information on hypernuclei, in particular 
by fitting to the optical potential of hyperons extracted from the data. Finally, the 
coupling of the $\phi$ meson to the $\Lambda$ baryon is reduced by 20\% from its SU(3) 
value in order to reproduce $\Lambda\Lambda$ bond energy data \citep{Ahn:2013poa}.

The EoS of neutron star matter and the M-R relation from the FSU2R model with inclusion 
of hyperons---dubbed as FSU2R(hyp) model---are plotted, respectively, in 
Figs.~\ref{fig:eosbeta} and~\ref{fig:MRplot}. As expected, due to the softening of the 
high-density EoS with hyperonic degrees of freedom (compare the FSU2R(hyp) and 
FSU2R(nuc) EoS's in Fig.~\ref{fig:eosbeta}), we obtain a reduction of the maximal 
neutron star mass below 2$M_{\odot}$ in the FSU2R(hyp) calculation. 
However, we may readjust the parameters of the nuclear model by stiffening 
the EoS of isospin-symmetric matter for densities above twice the saturation density, 
i.e., the region where hyperons set in, simultaneously preserving  
the properties of the previous EoS for the densities near saturation, which are 
important for finite nuclei and for stellar radii. The couplings of the hyperons 
to the different mesons can be determined as before. 
The parameters of the new interaction \citep{Tolos:2016hhl,Tolos:2017lgv}, denoted as 
FSU2H, are displayed in Table~\ref{t-parameters}, along with the predicted symmetry 
energy $E_{\rm sym}(n_0)$ at saturation density and its slope $L$, which are 
safely within current empirical and theoretical bounds 
(cf.\ Fig.~4 of \cite{Tolos:2016hhl} and Fig.~1 of \cite{Tolos:2017lgv}).  
The neutron star calculations with the FSU2H model with allowance for hyperons in the 
stellar core---FSU2H(hyp) model---successfully fulfill the 2$M_{\odot}$ mass limit 
with moderate radii for the star (see Fig.~\ref{fig:MRplot} and 
Table~\ref{tab:starprops}), while the base nuclear model FSU2H still reproduces the 
properties of nuclear matter and nuclei. In isospin-symmetric nuclear matter, FSU2H 
leads to a certain overpressure in the EoS at high densities \citep{Tolos:2016hhl}, but 
the EoS in neutron matter satisfies the constraints from HICs 
\citep{Danielewicz:2002pu}, as can be seen in Fig.~\ref{fig:eosbeta}.

We also draw in Fig.~\ref{fig:MRplot} the \mbox{M-R} relation 
predicted by FSU2H if one neglects hyperons---FSU2H(nuc) model. 
As expected, since the hyperonic FSU2H(hyp) EoS is softer than the 
FSU2H(nuc) EoS after hyperons appear (see Fig.~\ref{fig:eosbeta}), the neutron star 
calculations with FSU2H(nuc) lead to a higher maximum mass than FSU2H(hyp). In 
the next section, comparisons between cooling calculations performed with FSU2H(hyp) and FSU2H(nuc) will 
be used to discuss the influence of the occurrence of hyperons on neutron star cooling.

\vspace*{5mm}
\section{Cooling from low to high-mass neutron stars}
\label{sec:results}

Once the microscopic models for the EoS and the resulting properties of neutron stars have been discussed, we proceed to calculate the thermal evolution of such stars. We recall that the cooling of a neutron star is driven by neutrino emission from its interior, as well as photon emission from the surface. The equations that govern their cooling are those of thermal balance and of thermal energy transport \citep{2006NuPhA.777..497P,1999Weber..book,1996NuPhA.605..531S}, given by ($G = c = 1$) 
\begin{eqnarray}
  \frac{ \partial (l e^{2\Phi})}{\partial m}& = 
  &-\frac{1}{\rho \sqrt{1 - 2m/r}} \left( \epsilon_\nu 
    e^{2\Phi} + c_v \frac{\partial (T e^\Phi) }{\partial t} \right) \, , 
  \label{coeq1}  \\
  \frac{\partial (T e^\Phi)}{\partial m} &=& - 
  \frac{(l e^{\Phi})}{16 \pi^2 r^4 \kappa \rho \sqrt{1 - 2m/r}} 
  \label{coeq2} 
  \, .
\end{eqnarray}
The cooling of neutron stars depends on both micro and macroscopic ingredients, as can be seen 
in eqs.~(\ref{coeq1}) and (\ref{coeq2}), where all symbols have their usual meaning and the thermal variables are the neutrino emissivity $\epsilon_\nu(r,T)$, the thermal conductivity $\kappa(r,T)$, the specific heat $c_v(r,T)$, the luminosity  $l(r,t)$, and the temperature $T(r,t)$.

In addition to Eqs.~(\ref{coeq1}) and (\ref{coeq2}), one also needs a boundary condition connecting the surface temperature to that in the mantle \citep{Gudmundsson1982,Gudmundsson1983,Page2006}, as well as the condition of zero luminosity at the center in order to satisfy the vanishing heat flow at this point. 
In this study, we make use of all neutrino emissivities allowed for the EoS's and the corresponding compositions, that is, all processes involving nucleons and, when pertinent, hyperons, as well as the appropriate specific heat and thermal conductivity. A thorough review of such processes can be found in reference \cite{Yakovlev2000}.

\subsection{Cooling of neutron stars without nucleon pairing}

\begin{figure}[t]
\begin{center}
\includegraphics[width=1.1\columnwidth]{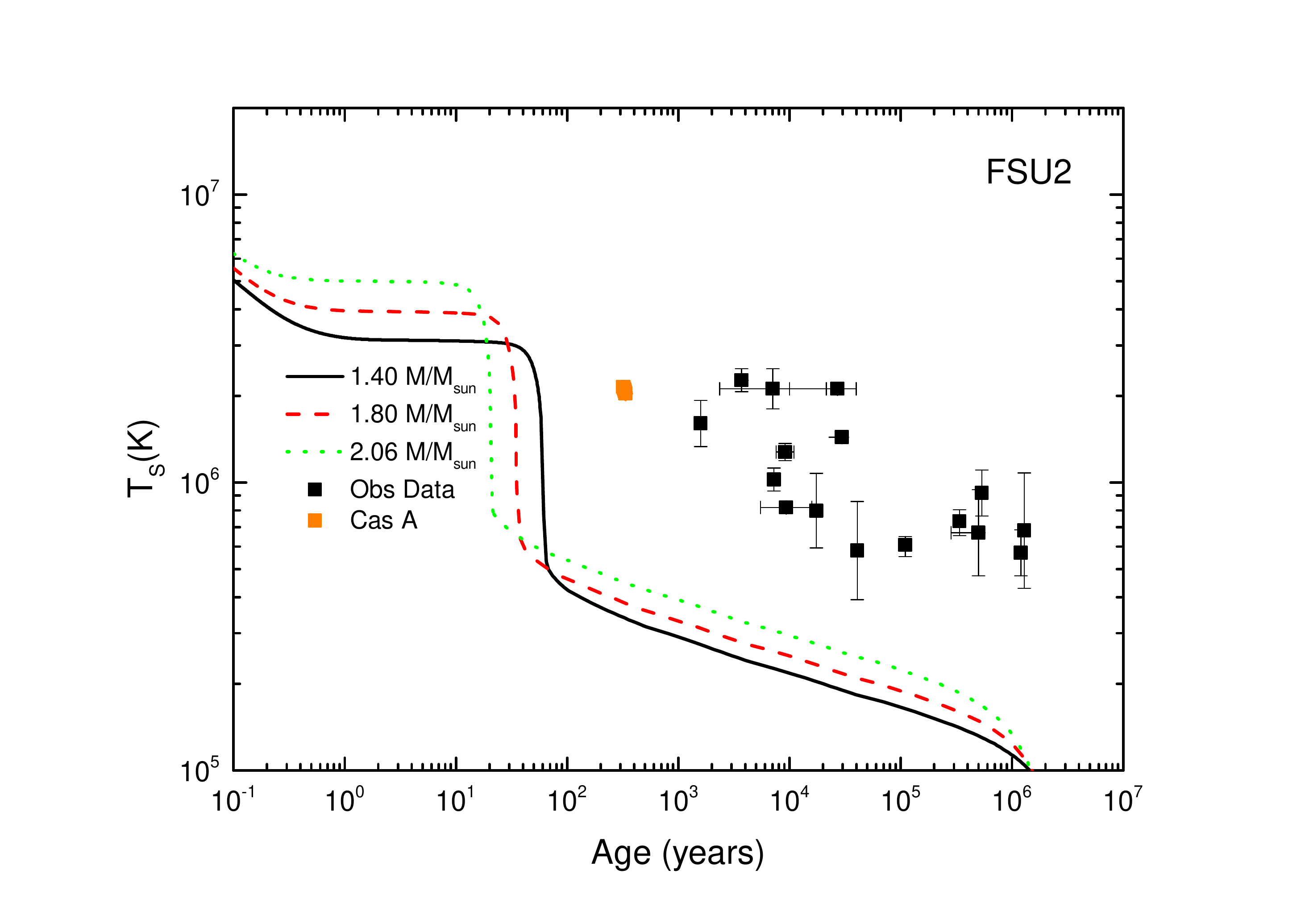}
\caption{Surface temperature as a function of the stellar age for different 
neutron star masses in the FSU2 model. Also shown are different observed thermal 
data.}
\label{fig:coolNOSF1}
\end{center}
\end{figure}

\begin{figure}[t]
\begin{center}
\includegraphics[width=1.1\columnwidth]{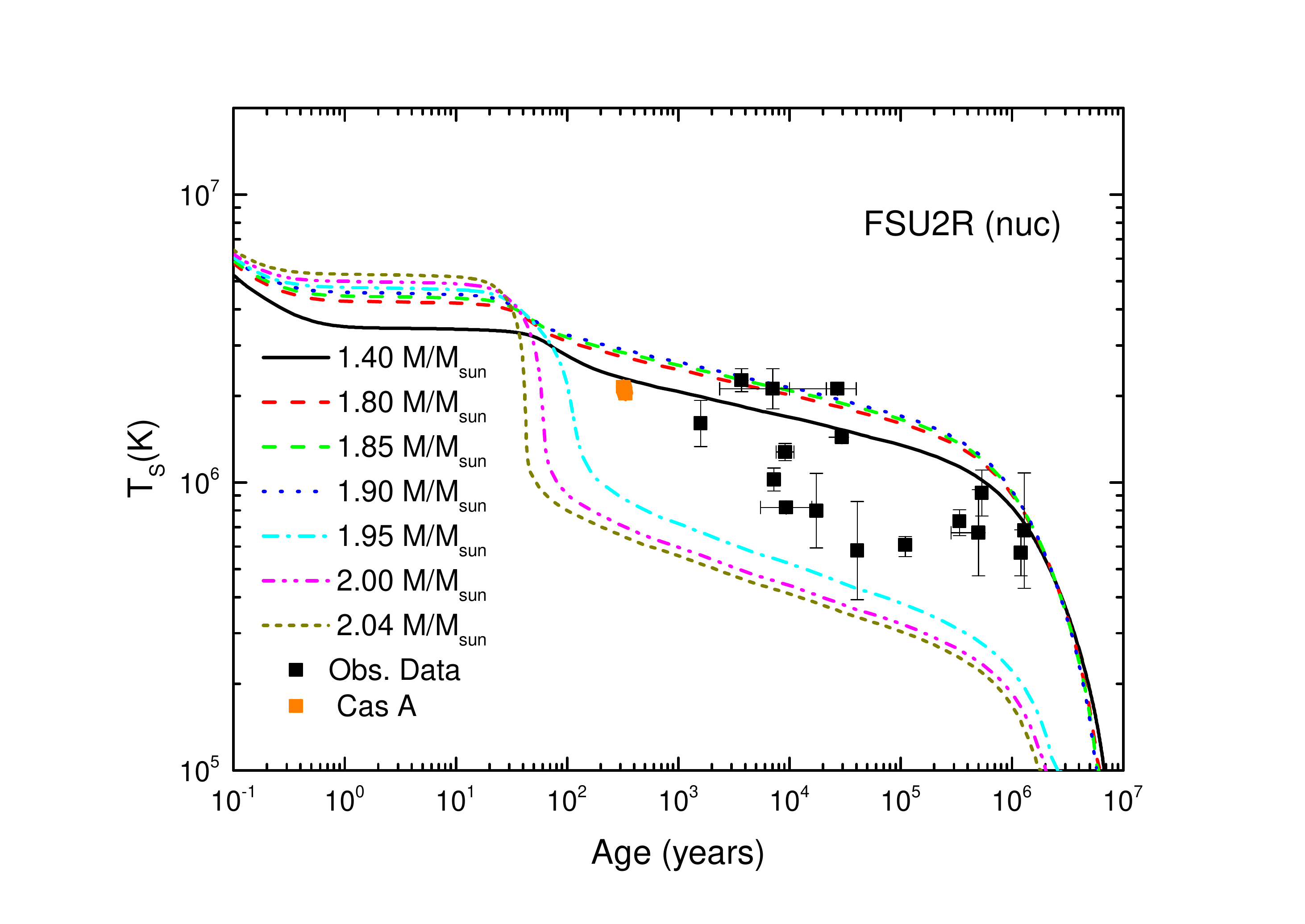}
\caption{Same as in Fig.~\ref{fig:coolNOSF1} but for the FSU2R (nuc) model. }
\label{fig:coolNOSF5}
\end{center}
\end{figure}

\begin{figure}[t]
\begin{center}
\includegraphics[width=1.1\columnwidth]{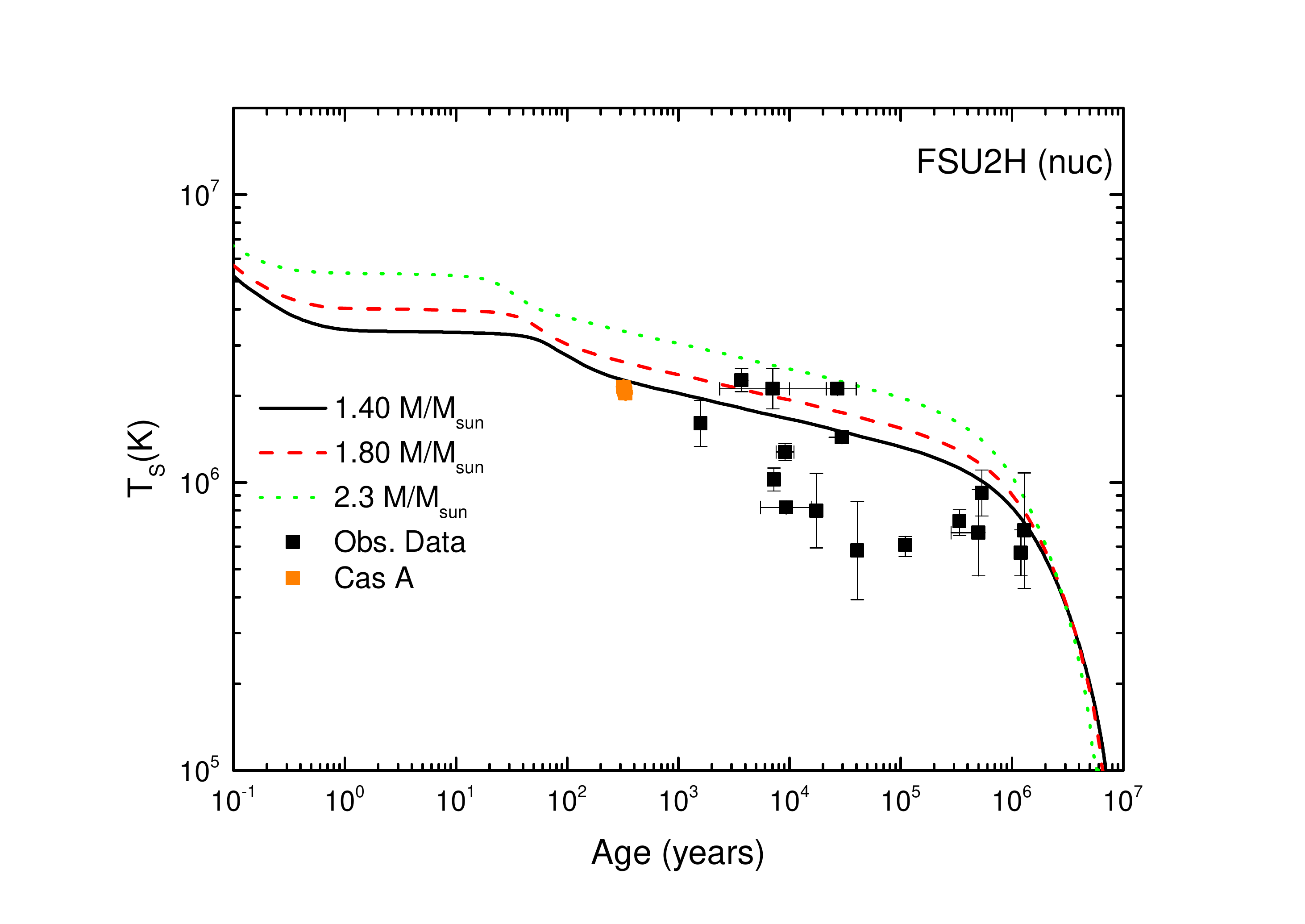}
\caption{Same as in Fig.~\ref{fig:coolNOSF1} but for the FSU2H (nuc) model. }
\label{fig:coolNOSF3}
\end{center}
\end{figure}

\begin{figure}[t]
\begin{center}
\includegraphics[width=1.1\columnwidth]{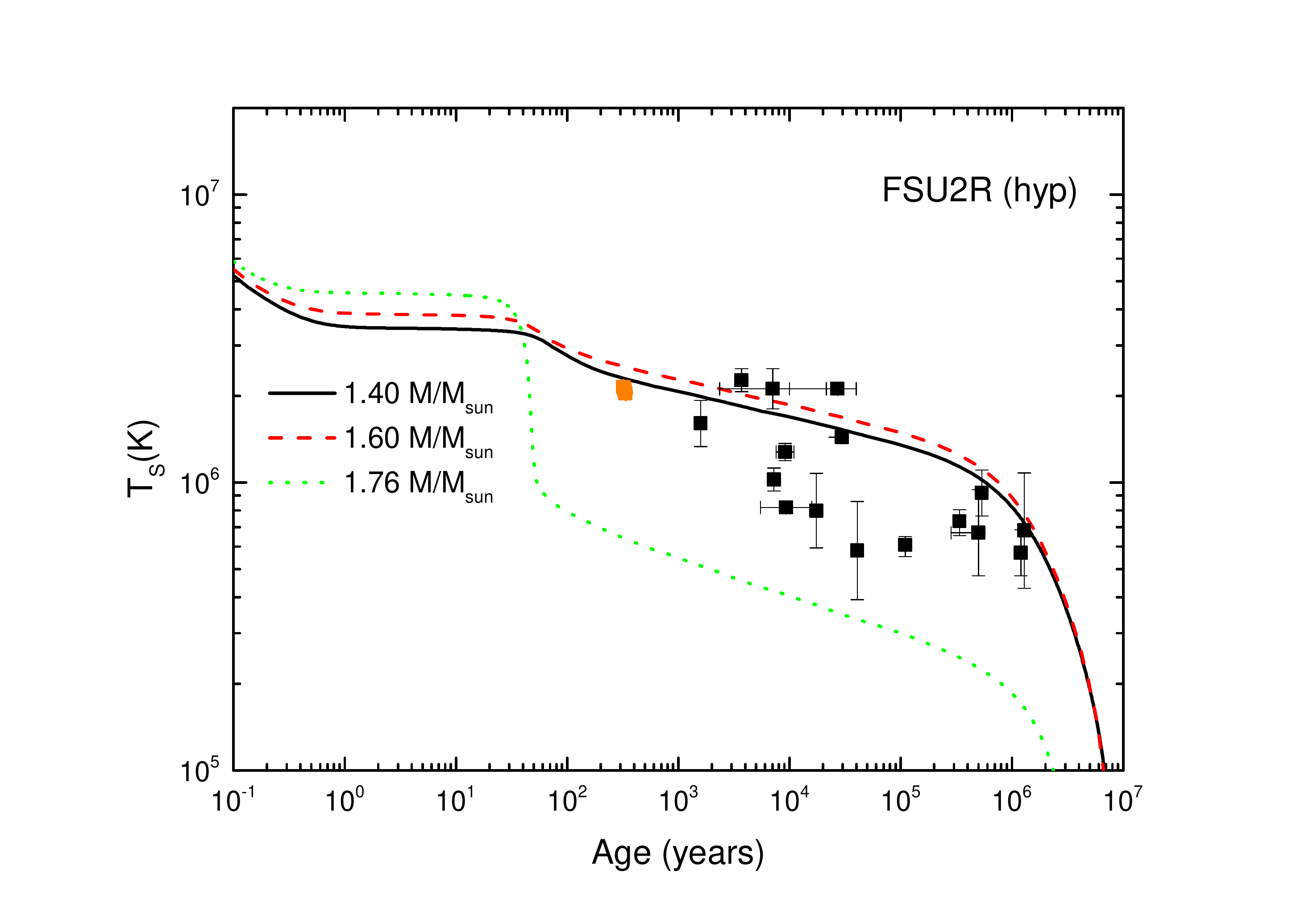}
\caption{Same as in Fig.~\ref{fig:coolNOSF1} but for the FSU2R (hyp) model. }
\label{fig:coolNOSF4}
\end{center}
\end{figure}

\begin{figure}[t]
\begin{center}
\includegraphics[width=1.1\columnwidth]{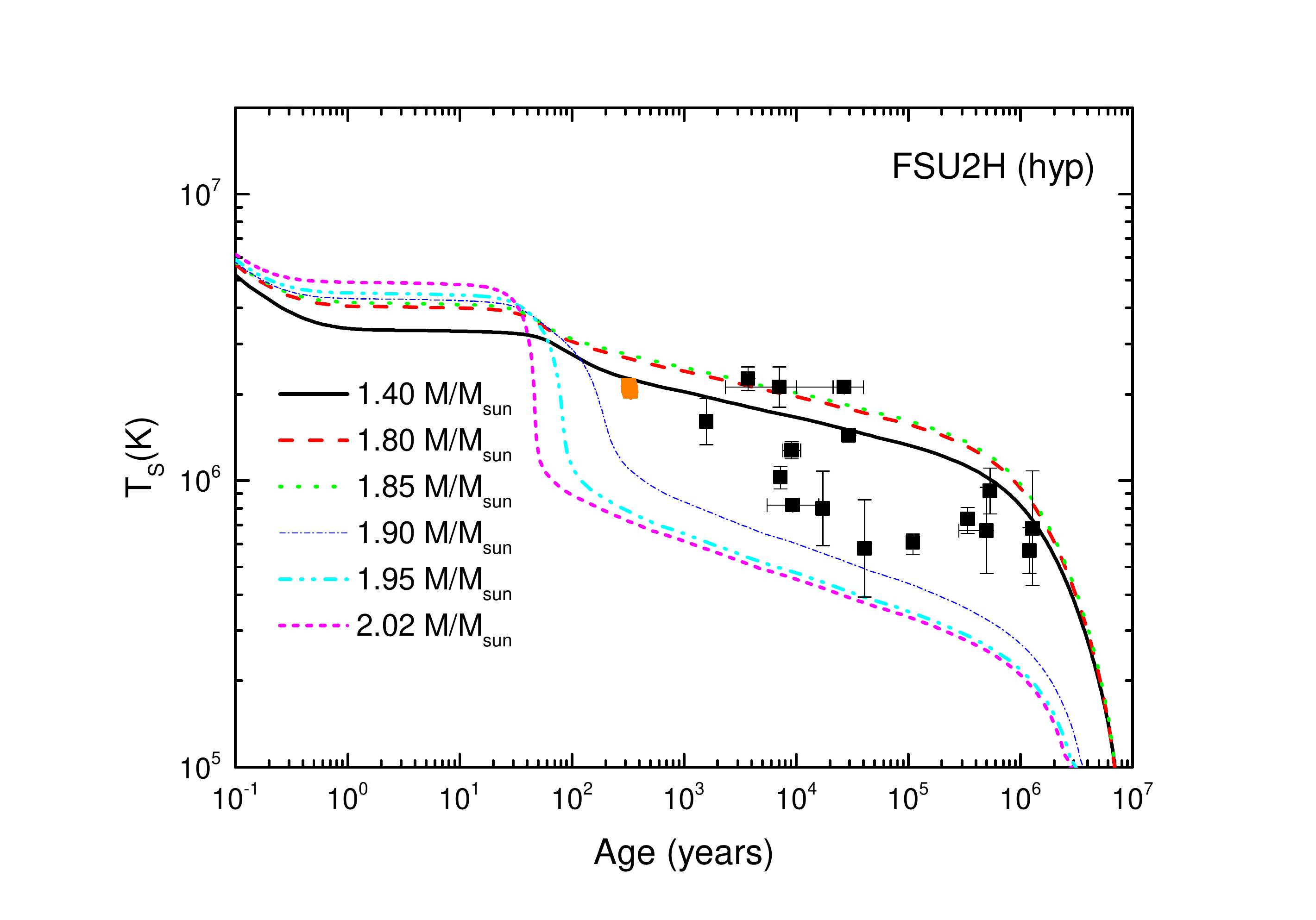}
\caption{Same as in Fig.~\ref{fig:coolNOSF1} but for the FSU2H (hyp) model. }
\label{fig:coolNOSF2}
\end{center}
\end{figure}

We first consider the thermal evolution of neutron stars without taking into account any sort of pairing 
(neither in the core nor in the crust). This is done in  Figs.~\ref{fig:coolNOSF1}--\ref{fig:coolNOSF2} with the ultimate goal 
of determining how the different models for the EoS describe the thermal behavior of neutron stars as ``benchmark testing''
before the inclusion of pairing.
However, these results should not be regarded as advocating for the absence of superfluidity/superconductivity in the star,
whose effect will be considered in Subsection \ref{sec:pairing}. In Figs.~\ref{fig:coolNOSF1}--\ref{fig:coolNOSF2}
we depict several cooling curves for low, medium and high neutron star masses for each model, in order to investigate the cooling behavior for different central density regimes.
The figures also display the observed surface temperature versus the age of a set of prominent neutron stars, including that of the remnant in Cassiopeia A \citep{Beloin:2016zop,SafiHarb2008,Zavlin1999,Pavlov2002,Mereghetti1996,Zavlin2007,Pavlov2001,Gotthelf2002,McGowan2004,Klochkov2015,McGowan2003,McGowan2006,Possenti1996,Halpern1997,Pons2002,Burwitz2003,Kaplan2003,Zavlin,Ho2015}.

\begin{table*}[t]
\begin{center}
\begin{tabular}{| c c c c c c c c c|}
\hline
Models & DU threshold & hyp DU threshold&  $1.4 M_\odot$   &   $1.4 M_\odot$  &$1.76 M_\odot$  & $ 1.76 M_\odot$   &$ 2.0 M_\odot$ & $2.0 M_\odot$ \\
&(fm$^{-3}$) &(fm$^{-3}$) & $n_c$ (fm$^{-3}$) & cooling & $n_c$ (fm$^{-3}$) & cooling & $n_c$ (fm$^{-3}$)  & cooling  \\
\hline
FSU2 (nuc) \ \ & 0.21 & | & 0.35 & fast & 0.47 & fast & 0.64 & fast \\
FSU2R (nuc) & 0.61 & |    & 0.39 & slow & 0.51 & slow & 0.72 & fast \\
FSU2H (nuc) & {0.61} & |    & 0.34 & slow & 0.39 & slow & 0.45 &  slow \\
FSU2R (hyp) & 0.57 & 0.37 & 0.40 & slow & 0.87 & fast & | &  | \\
FSU2H (hyp) & 0.52 & 0.34 & 0.34 & slow & 0.44 & slow & 0.71 &  fast \\
\hline
 \end{tabular}
\end{center}
\caption{Thermal behavior of the different models studied. The DU threshold indicates the density at which the URCA process of nucleons becomes effective. Similarly, the hyp DU threshold indicates the density at which the hyperonic DU processes, given by the $\Lambda$ particle, begin to act. Also shown is the central density $n_c$ for three selected neutron star masses, as well as whether such stars exhibit slow or fast cooling.}
\label{table:cool1}
\end{table*}

The cooling behavior from each microscopic model is summarized in Table~\ref{table:cool1} for low-mass to high-mass neutron stars. 
When DU reactions of neutrino production are allowed according to the model, they lead to an enhanced cooling of the star. 
We summarize here the DU processes that may take place inside the neutron star:
\begin{eqnarray}
n \rightarrow p + e^- +\bar{\nu}, \nonumber \\
\Lambda \rightarrow p + e^- +\bar{\nu}, \nonumber\\
\Sigma^- \rightarrow n + e^- +\bar{\nu}, \nonumber\\
\Sigma^- \rightarrow \Lambda + e^- +\bar{\nu}, \nonumber\\
\Sigma^- \rightarrow \Sigma^0+ e^- +\bar{\nu}, \nonumber \\
\Xi^- \rightarrow \Lambda+ e^- +\bar{\nu}, \nonumber\\
\Xi^- \rightarrow \Sigma^0+ e^- +\bar{\nu}, \nonumber\\
\Xi^0 \rightarrow \Sigma^+ + e^- +\bar{\nu}, \nonumber\\
\Xi^- \rightarrow \Xi^0 + e^- +\bar{\nu} . \nonumber 
\end{eqnarray}
We stress that the presence of the particles is not enough for such processes to take place, and one must also account for momentum conservation, as previously discussed. We 
also recall that all inverse reactions also take place as to maintain (on average) chemical equilibrium.
In Table~\ref{table:cool1} we indicate the density threshold for the nucleonic DU process and, when applicable, the threshold for the hyperonic DU processes, determined by the $\Lambda$ particle as the first hyperon to appear. We also show the central density $n_c$ for three selected neutron star masses, as well as whether such stars exhibit slow or fast cooling scenarios. The information supplied in Table~\ref{table:cool1} will help to understand the thermal evolution of neutron stars, displayed in Figs.~\ref{fig:coolNOSF1} to \ref{fig:coolNOSF2} for the different models and discussed in the following:

{\bf FSU2} (Fig.~\ref{fig:coolNOSF1}): All neutron stars whose microscopic composition is described by this model allow for pervasive DU process, even for low-mass stars with low central densities of around 2$n_0$, leading to a cooling which is too fast  in comparison with the observed data. 

{\bf FSU2R(nuc)} (Fig.~\ref{fig:coolNOSF5}): For this model most stars (up to $M\sim 1.90$$M_{\odot}$) exhibit slow cooling. This model could, in principle, explain most of the observed data. However, it would mean that all colder stars with $T \lesssim 10^6$ K have masses higher than $1.9 M_\odot$, which seems unlikely. 

{\bf FSU2H(nuc)} (Fig.~\ref{fig:coolNOSF3}): For this model the DU process is absent in all stars studied, from low-mass stars with lower central densities to high-mass stars with higher central densities. Therefore, all stars  exhibit slow cooling, which allows for agreement with some of the observed data but fails to explain most of the observed cold stars. 

{\bf FSU2R(hyp)} (Fig.~\ref{fig:coolNOSF4}): The results of this model exhibit a relatively similar pattern to that of the FSU2R(nuc) model, with the exception that the highest mass possible in this model is {$1.765 M_\odot$}, and that only stars with masses similar to that one exhibit fast cooling. 

{\bf FSU2H(hyp)} (Fig.~\ref{fig:coolNOSF2}): {This particular model shows the most promising results, with stars with lower masses exhibiting slow cooling (without DU) and higher masses displaying fast cooling (with DU), while intermediate masses show a behavior in between these extremes. This model explains most of the observed data, including a reasonable agreement with Cas A, without the need of resorting to extensive pairing. Nevertheless, in this case, a large portion of the observational data corresponds to stars with masses between the restricted range of 1.85 and 1.90 $M_\odot$. This issue becomes less problematic when nucleonic pairing is included, as will be shown in the following subsection.}

The different cooling pattern for each model can be understood from the microscopic differences between the models. 
We start by analyzing the case of low-mass stars of $1.4 M_{\odot}$ with central densities around twice nuclear saturation 
density. The different cooling pattern of a $1.4 M_{\odot}$ star observed for
FSU2 (fast cooling) and FSU2R(nuc) (slow cooling) is mainly due to the different density dependence of the symmetry energy 
around saturation in these models and, hence, to the different value of the symmetry energy slope parameter ($L$), which can be read in 
Table~\ref{t-parameters} ($L\approx113$ MeV in FSU2 and $L\approx47$ MeV in FSU2R). The 
larger the value of $L$ is, the more protons are produced and, thus, the DU process appears at lower densities, making the
cooling of the star more efficient. Indeed, even if a $M =1.4$$M_{\odot}$ star in the FSU2 
model has a smaller central density than in the 
FSU2R(nuc) model, the cooling is faster in FSU2 as the DU threshold for FSU2 is at a much smaller density, as seen in 
Table~\ref{table:cool1}. This conclusion is corroborated by the cooling behavior of 
FSU2R(nuc) and FSU2H(nuc), which show a similar qualitatively slow cooling for neutron 
stars with $M \sim 1.4 M_{\odot}$, as both models have an alike $L$ value (cf.\ 
Table~\ref{t-parameters}). Therefore, one finds that in low-mass stars, where  
the central density does not go much above  2$n_0$, 
large stellar radii (stiff nuclear symmetry energy near $n_0$) are associated with fast 
cooling, whereas small stellar radii (soft nuclear symmetry energy near $n_0$) imply 
slow cooling, as also seen in \cite{Dexheimer:2015qha}  in the framework of the Chiral Mean Field Model.

As for high-mass stars ($M \sim 1.8$--$2 M_\odot$), we find that the 
different behaviors exhibited by the cooling curves of FSU2 (fast), FSU2R(nuc) (fast) and 
FSU2H(nuc) (slow) are correlated with the different values of the central densities in these 
stars, as seen in Table~\ref{table:cool1}. This is understood by considering the fact 
that the FSU2H(nuc) model produces a stiffer EoS in the region of high densities than 
the FSU2 and FSU2R(nuc) models (see Fig.~\ref{fig:eosbeta}). Therefore, for the same 
heavy stellar mass, neutron stars obtained within the FSU2H(nuc) model have much lower 
central densities than in FSU2 and FSU2R(nuc), making the DU process less efficient and, 
thus, leading to a slower cooling. For these high stellar masses, the densities reached are 
much higher than saturation density and, therefore, the slope of the symmetry energy at saturation is 
not determinant for the cooling behavior.

\begin{figure}[th]
\begin{center}
\includegraphics[width=0.9\columnwidth]{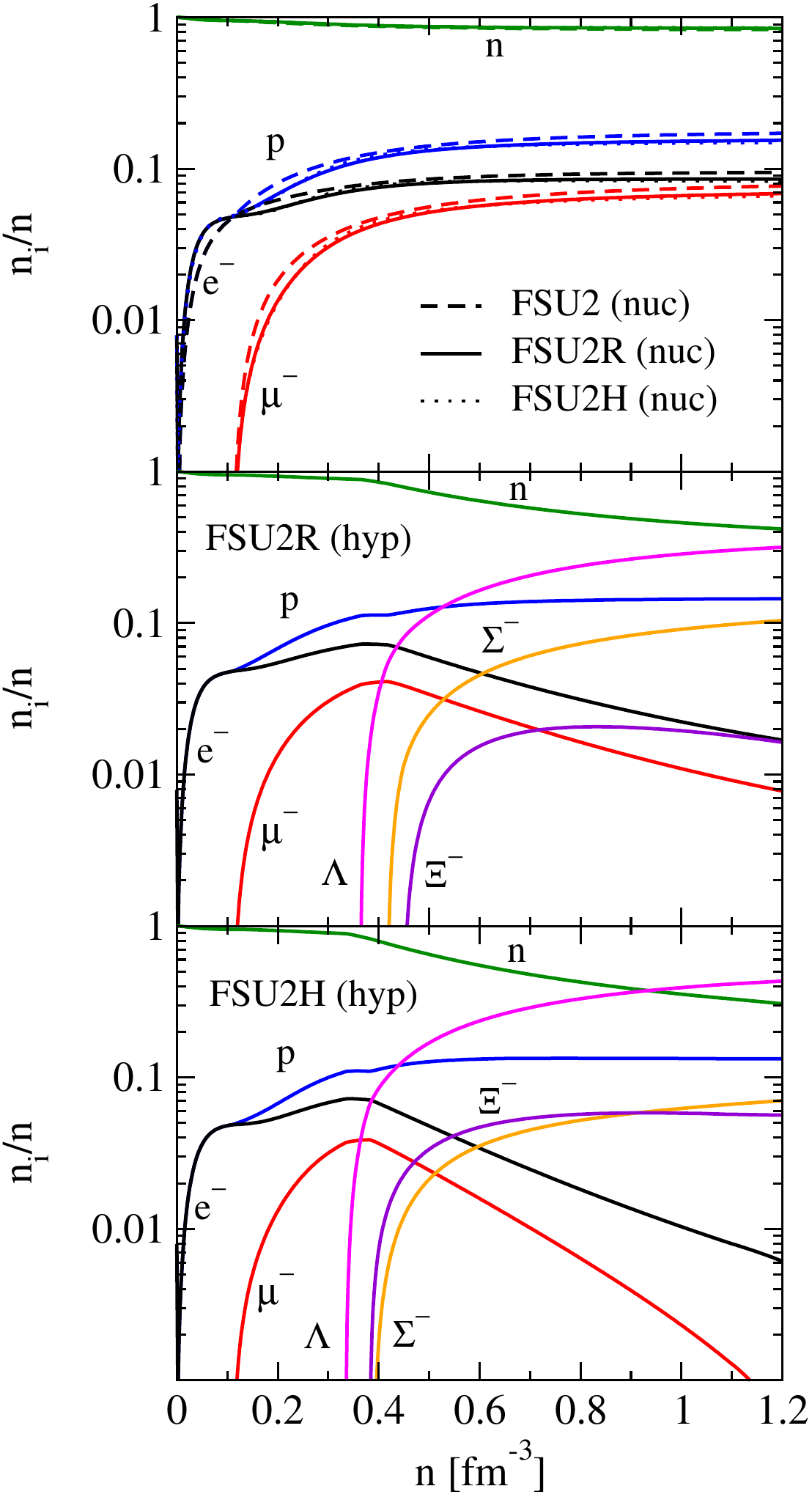}
\caption{Particle fractions as functions of the baryonic density for the models discussed in the text.}
\label{fig:fractions}
\end{center}
\end{figure}

With regards to the models that include hyperons, as their EoS's are softer than their nucleonic counterparts, they produce stars with higher central density values, which may then overcome the DU threshold, as one can clearly see in Table~\ref{table:cool1}. As a consequence, stars with $M \sim 1.76 M_{\odot}$ that cooled more slowly without hyperons in the FSU2R(nuc) model, change to a faster cooling pattern in the presence of hyperons, as seen for FSU2R(hyp).
We also observe in Table~\ref{table:cool1} that hyperonic models activate
 the DU process at similar or even lower densities than the models without hyperons. This is mainly due to the fact that the presence of hyperons (mostly $\Lambda$ particles) reduces the neutron fraction at a given baryon number density and, consequently, the DU constraint 
$\vec{k}_{Fn}=\vec{k}_{Fp}+\vec{k}_{Fe}$, with ${k}_{Fn}$, ${k}_{Fp}$ and ${k}_{Fe}$ 
being the Fermi momenta of the neutron, proton and electron, respectively, can be 
fulfilled at a lower density. This is seen in Fig.~\ref{fig:fractions}, where the 
particle fractions are shown as functions of the neutron star matter density, for the 
various interaction models explored in this work. It is clear that the appearance of the 
$\Lambda$ hyperons between 0.3 and 0.4 fm$^{-3}$ for the FSU2R(hyp) and FSU2H(hyp) models 
(middle and lower panels) induces a substantial decrease in the neutron fraction 
compared to the purely nucleonic models collected in the upper panel.
In summary, the presence of hyperons in the cores of medium to heavy mass stars speeds up their cooling pattern. 
When comparing the cooling curves of the FSU2R(hyp) and FSU2H(hyp) 
models, we notice that higher stellar masses are needed in the case of FSU2H(hyp) to reach a fast 
cooling behavior. This is due to the fact that the EoS of the FSU2R(hyp) model is softer 
and the central densities achieved are larger compared to the ones for FSU2H(hyp),  
hence the DU threshold is overcome more easily, even if it appears at slightly higher densities. 
Finally, it may be noticed that the cooling pattern seen in Fig.~\ref{fig:coolNOSF2} for FSU2H(hyp) is similar to that in Fig.~\ref{fig:coolNOSF5} for FSU2R(nuc), 
so that it could argued that it is unnecessary to resort to a hyperonic model such as FSU2H(hyp) to explain most of the observed cooling data. 
However, purely nucleonic models are forcedly omitting the presence of hyperons, which are physically allowed at intermediate densities when the appropriate chemical equilibrium conditions are fulfilled.

\subsection{Neutron superfluidity and proton superconductivity effects on the cooling of neutron stars}
\label{sec:pairing}

We now turn our attention to investigate the neutron superfluidity and proton superconductivity on the cooling of neutron stars. Pairing effects have been considered a key factor for the thermal evolution of neutron stars \citep{Beloin:2016zop,Yakovlev2000,Gnedin2008,Weber2009a,Page2004,Weber2007}, as well as of extreme importance to find agreement between the theoretical models and the observed data, particularly for Cas A \citep{Page2011a,Ho2009,Yang2011,Heinke2010,Ho2015,Shternin2011a,Blaschke2012,Ho2009,Negreiros2013}. 

As indicated before, one of the major cooling channels in neutron stars, especially during the first $\sim$$10^3$~years, is neutrino production reactions. Most of these reactions involve baryons, and chief among those is the DU process, which, if present, is the leading cooling mechanism in neutron stars. The introduction of a superfluidity (conductivity) gap in the energy spectrum of such baryons reduces the reaction rates. One notes that the reduction factor depends on the temperature and its relation to the corresponding superfluidity (conductivity) critical temperature (or the gap), and leads to a sharp drop of neutrino emissivity after the matter temperature drops below the pairing critical temperature.  The calculation of the reduction factor for each neutrino emission process is a complicated procedure, which can be obtained by the study of the phase-space of the emission processes (see \cite{Yakovlev2000} for a comprehensive calculation of such factors).

\begin{figure}[t]
\begin{center}
\includegraphics[width=\columnwidth]{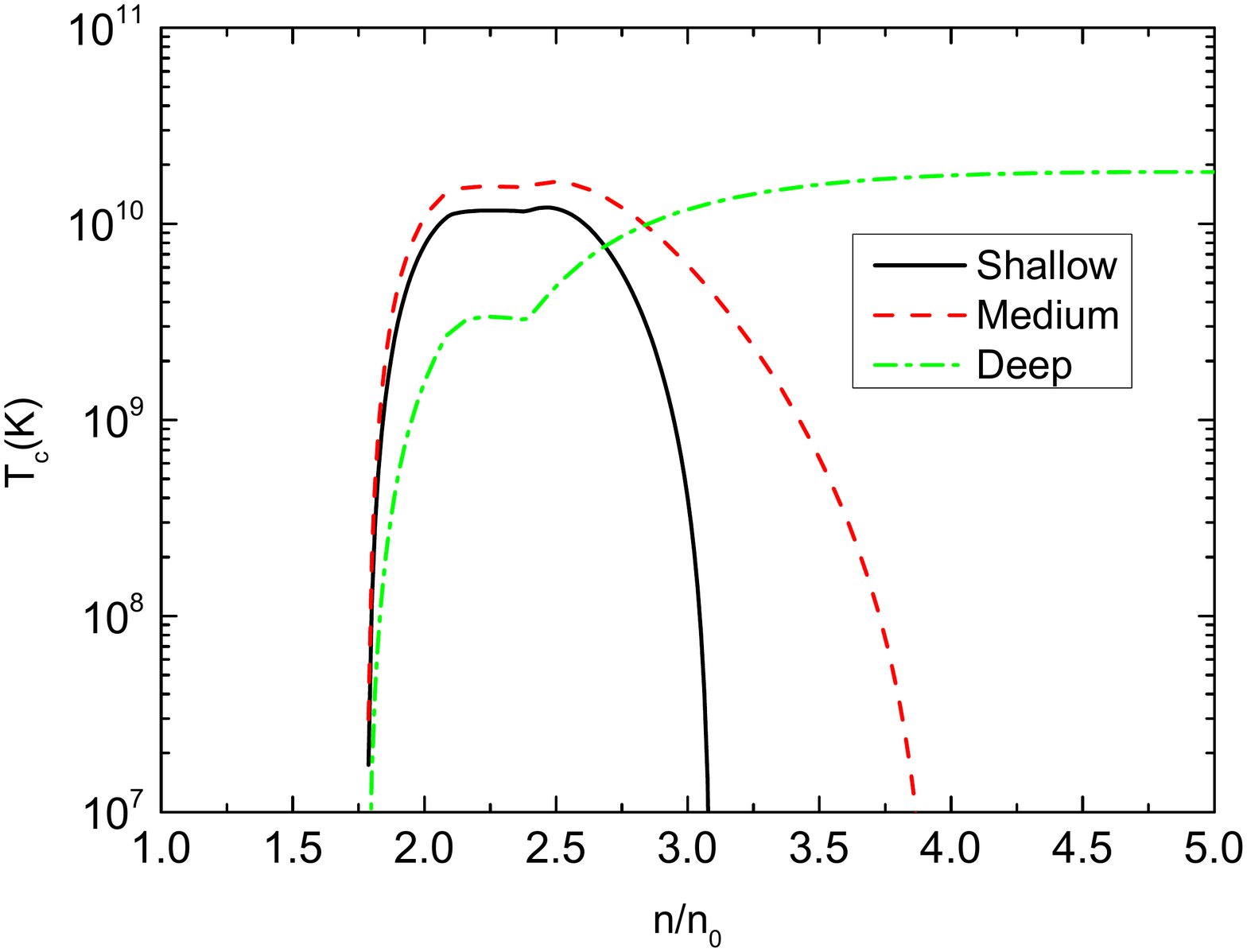}
\caption{Critical temperature for proton singlet ($^1S_0$) pairing as a function of normalized number baryon density for the three pairing scenarios studied. }
\label{fig:pSC}
\end{center}
\end{figure}

There is, however, a great deal of uncertainties regarding proton and neutron pairing in neutron stars \citep{Yakovlev2000,Page2004,2006NuPhA.777..497P}, especially at high densities, not to mention the possibility of pairing among hyperons, which is even more uncertain. {The inclusion of microscopic calculations for pairing is a challenging task, especially for proton superconductivity. For that reason we have chosen to follow a phenomenological approach like that of \cite{Kaminker:2001ag}, which gives us the flexibility to probe different scenarios of proton pairing.  Such an approach was used in the work of \cite{Shternin2011a} and others \citep{2004AstL...30..759G, 2005NuPhA.752..590Y, Gusakov2005}. Also the work of \cite{Beloin:2016zop} uses a similar phenomenological approach for pairing. 
 Here we explore three different scenarios for proton pairing that reflect three assumptions: } a) shallow proton pairing (limited to densities up to 2--3\,$n_0$), b) medium proton pairing (extending up to 4\,$n_0$), and c) deep proton pairing (extending beyond 5\,$n_0$, deep in the inner core). In this manner, we can estimate how extensive the proton pairing must be so as to allow for a good comparison with the observed data. 
In Fig.~\ref{fig:pSC} we show the critical temperature for proton singlet ($^1S_0$) pairing as a function of density for the three different scenarios. 
{We note that we have used a maximum critical temperature of the order of $10^{10}$~K that is above certain parameterizations used in the literature 
such as \cite{Page2004} ($T_{\rm c,p}^{\rm max} \lesssim 6\times 10^9$~K) and \cite{Kaminker:2001ag} ($T_{\rm c,p}^{\rm max} \sim 5 \times 10^9$~K),
and closer to \cite{Gusakov2005} ($T_{\rm c,p}^{\rm max} \sim 7 \times 10^9$~K) and \cite{Beloin:2016zop} ($T_{\rm c,p}^{\rm max} = 7.59^{+2.48}_{-5.81}\times 10^9$~K).
This choice, however, was incidental, as we have used a set of parameters for proton superconductivity that lead to such a high critical temperature. Nevertheless, such a choice makes little difference in the long-term behavior of the cooling of the star, as we will see later.}

\begin{figure}[t]
\begin{center}
\includegraphics[width=0.38\textwidth]{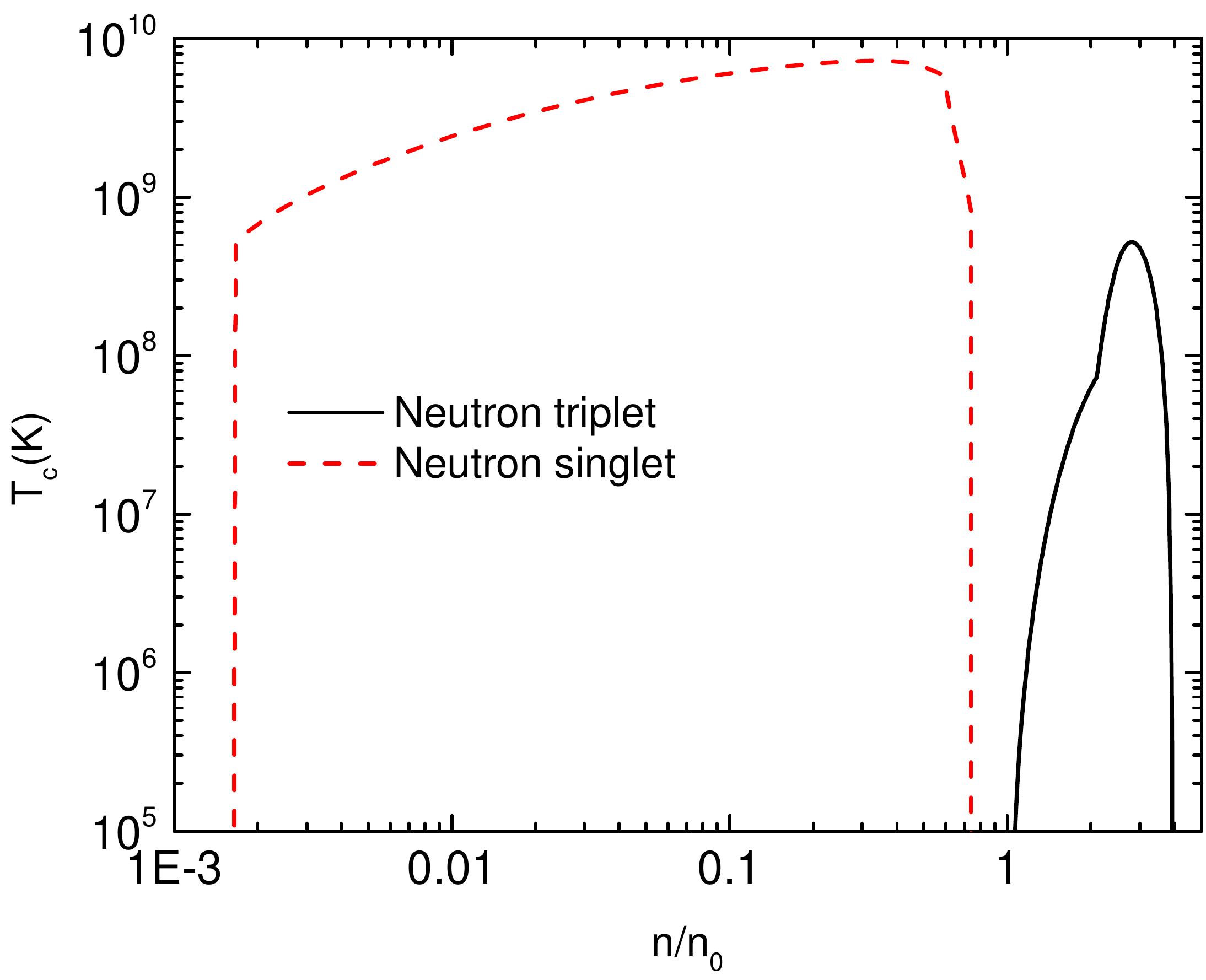}
\caption{Critical temperature for neutron pairing (singlet and triplet) as a function of normalized number baryon density. }
\label{fig:nSF}
\end{center}
\end{figure}

As for neutron pairing, somewhat less uncertain than that of protons (particularly for the neutron singlet pairing in the crust), we chose the standard {phenomenological}  approach: we allow for extensive neutron singlet ($^1S_0$) pairing at subnuclear densities (crust) and for a limited neutron triplet ($^3P_2$) pairing in the core, with a maximum critical temperature $\sim 5\times 10^8$ K,  similarly to the pairing used to explain the observed temperature of Cas A in \cite{Page2011a}. The critical temperature of both neutron  singlet and triplet pairings as a function of density can be seen in Fig.~\ref{fig:nSF}.

Apart from the suppression of neutrino emission reactions that involve paired particles, a consequence of pairing is the appearance of a new transient neutrino emission process, commonly known as pair breaking-formation (PBF) process. This process can be represented as  $b  \rightarrow b + \nu +\bar \nu$, with $b$ denoting paired baryons. It can be also understood as the annihilation of two quasi-baryons with similar anti-parallel momenta into a neutrino pair \citep{1976ApJ...205..541F,Voskresensky:1987hm,Kaminker:1999ez,Leinson:2006gf,Sedrakian:2006ys,Kolomeitsev:2008mc,Kolomeitsev:2010hr,Steiner:2008qz}. This process, also comprehensively described in \cite{Yakovlev2000}, is transient, reaching a maximum near the superfluidity (conductivity) onset and decreasing afterwards.

\begin{figure}[t]

\hspace*{-1cm}\includegraphics[width=1.25\columnwidth]{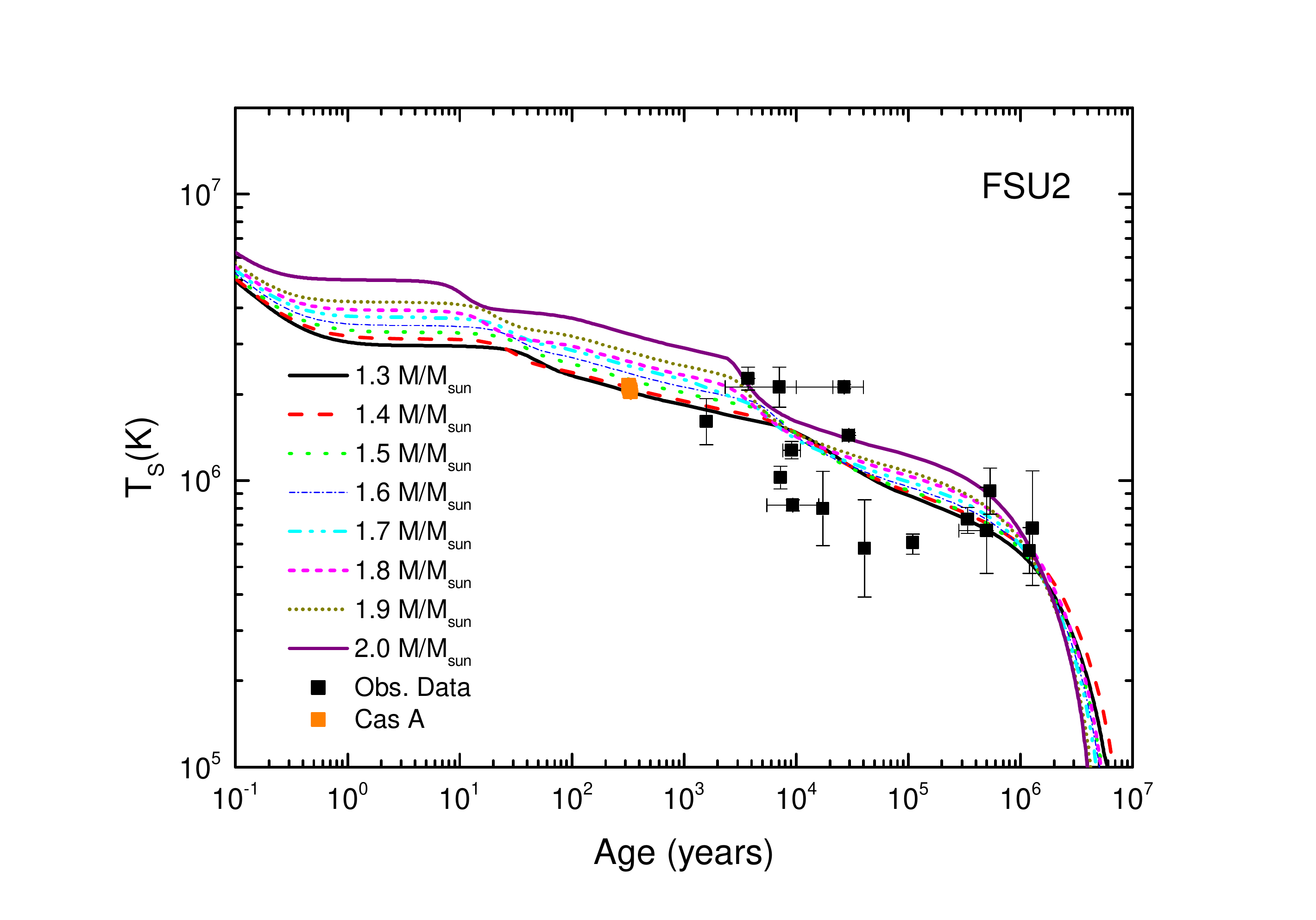}
\begin{center}
\caption{Surface temperature as a function of the stellar age for different neutron star masses in the FSU2 model subjected to deep proton pairing as well as neutron pairing.}
\label{fig:FSU2DEEP}
\end{center}
\end{figure}

For the analysis of pairing effects on the cooling of neutron stars, we concentrate our study on the two most relevant cases, the FSU2R(nuc) and the FSU2H(hyp) models \citep{Tolos:2016hhl,Tolos:2017lgv}, which are the ones that best reproduce the observed data on cooling in the previous section. We also show the predictions from the FSU2(nuc) model \citep{Chen:2014sca} for comparison. 
The main features of the results are as follows:

{\bf FSU2} (Fig.~\ref{fig:FSU2DEEP}): The original FSU2 model, as discussed before, has a small DU threshold, leading to fast cooling for all stars studied. The inclusion of shallow and medium proton pairing (in addition to neutron pairing, common to all simulations) is ineffective in slowing down the thermal evolution. Although part of the DU (among other processes) is suppressed, there is still a relatively large region at high densities in which the DU takes place, thus leading to a cooling behavior very similar to that of Fig.~\ref{fig:coolNOSF1}. The deep proton pairing, on the other hand, is effective in slowing the cooling, and leads to a slow cooling behavior, as shown in Fig.~\ref{fig:FSU2DEEP}. While the first ``knee" in the cooling curves, which usually happens between 50--150 years, is associated with the core-crust thermal coupling, the second ``knee"  around $\sim 5 \times 10^3$ years is linked to the onset of neutron pairing and the subsequent production of neutrinos coming from the PBF process.

\begin{figure}[t]
\hspace*{-1cm}\includegraphics[width=1.25\columnwidth]{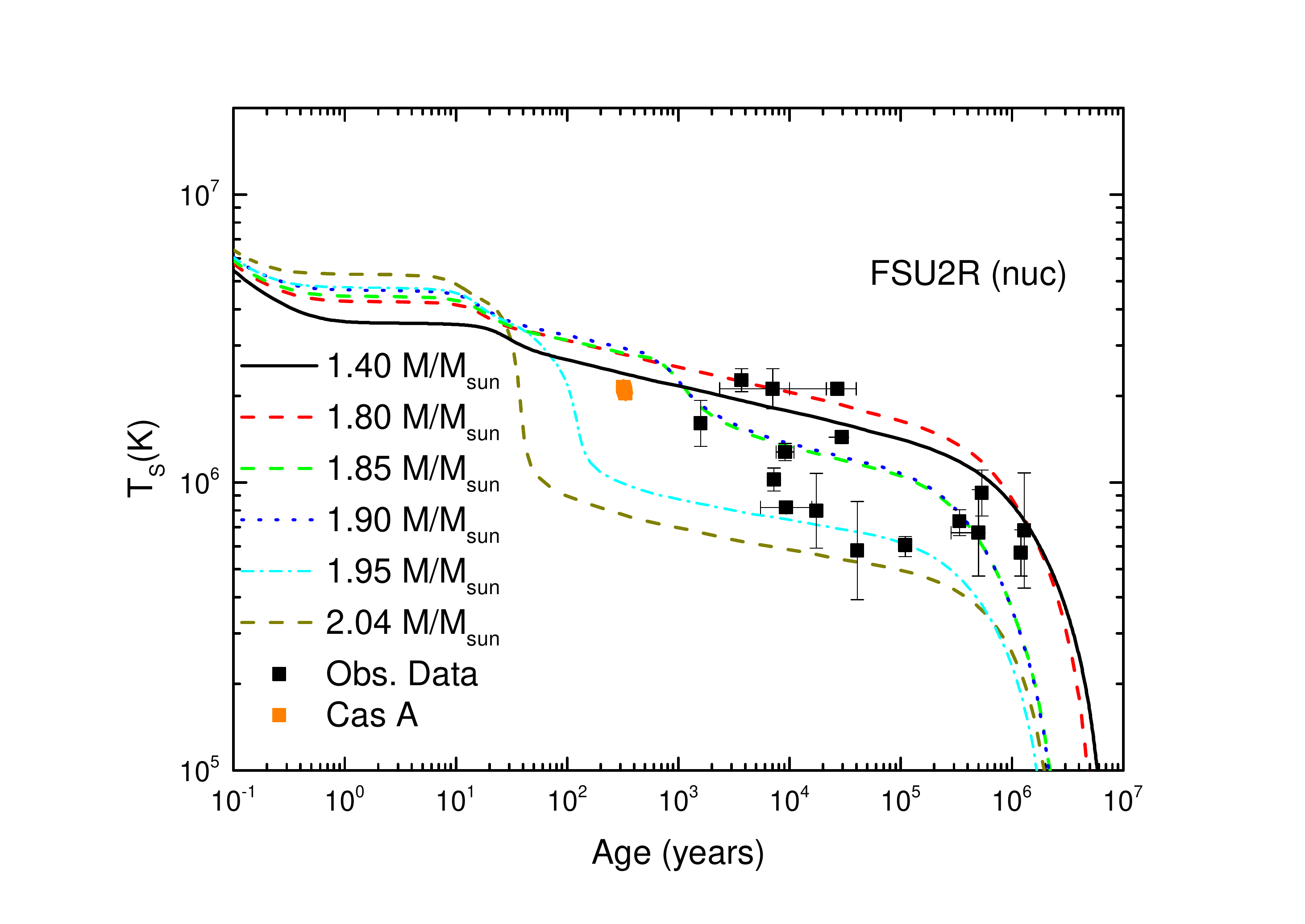}
\begin{center}
\caption{Same as Fig.~\ref{fig:FSU2DEEP} but for the FSU2R(nuc) model subjected to medium proton pairing as well as neutron pairing.}
\label{fig:FSU2RDEEP}
\end{center}
\end{figure}

{\bf FSU2R(nuc)} (Fig.~\ref{fig:FSU2RDEEP}): As previously seen, this model exhibits slow cooling only for stars below $1.9 M_\odot$ as the DU appears at high densities of $n_0 \sim 0.61$ fm$^{-3}$. Shallow to medium proton pairing, plus neutron pairing, allows for a satisfactory agreement with data, with the caveat that all observations with $T \lesssim 10^6$ K would need to be stars of relatively high mass (above $1.9 M_{\odot}$) within this model, as seen in Fig.~\ref{fig:FSU2RDEEP}. Deep proton pairing leads to the complete absence of DU processes, thus leading to a slow cooling scenario for neutron stars of all masses.

\begin{figure}[t]
\hspace*{-1cm}\includegraphics[width=1.25\columnwidth]{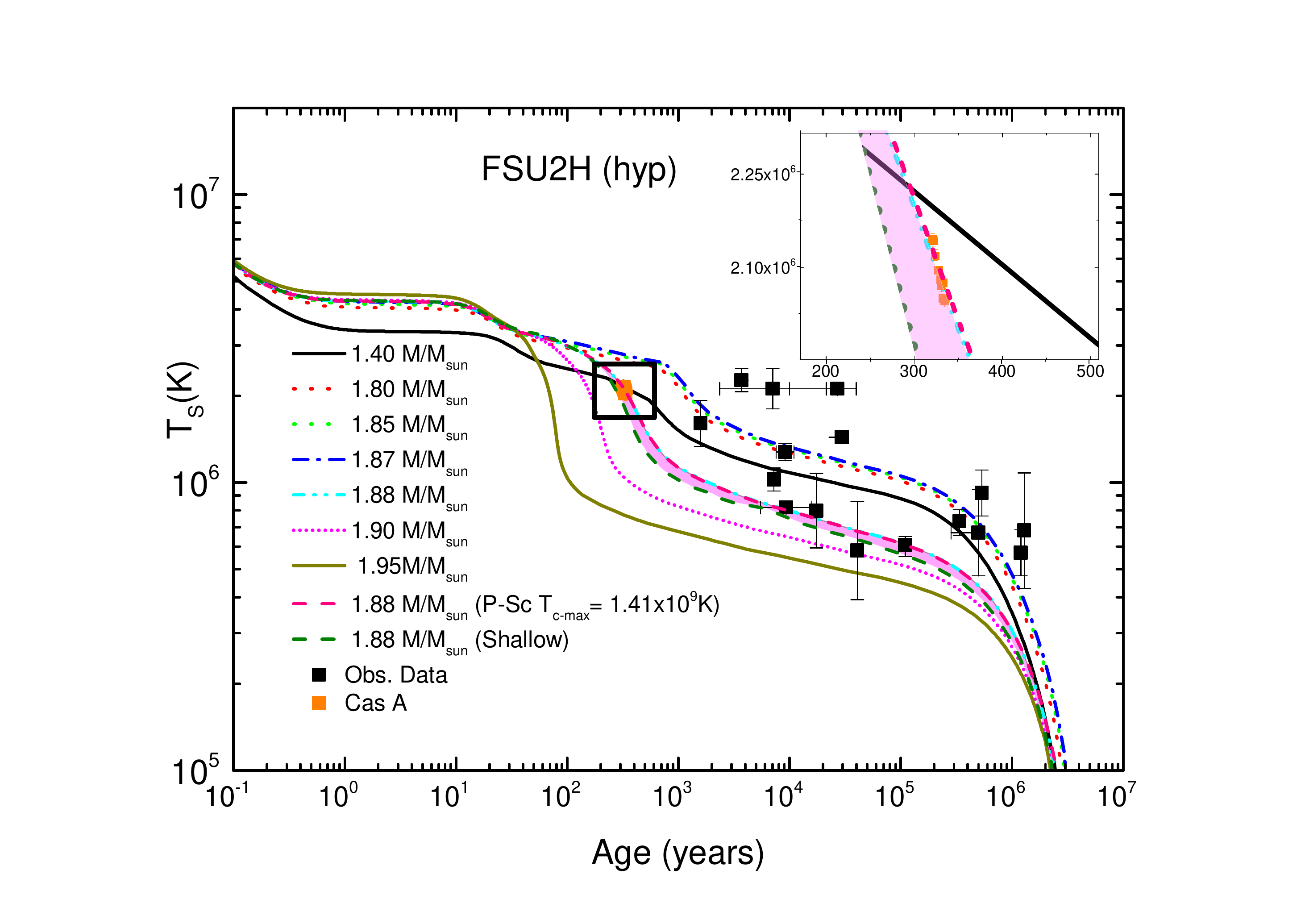}
\begin{center}
\caption{Same as Fig.~\ref{fig:FSU2DEEP} but for the FSU2H(hyp) model subjected to medium proton pairing as well as neutron pairing.
For a $1.88 M_{\odot}$ star we also show, on the one hand, the result from assuming a maximum critical temperature
of $1.41 \times 10^9$~K for proton superconductivity, and, on the other hand, the result from using shallow
proton pairing. The difference between the medium and shallow proton pairing scenarios is highlighted as the pink band.
The inset in the upper-right corner of the figure is a magnification of the region around Cas A.}
\label{fig:FSU2HBEST}
\end{center}
\end{figure}

{\bf FSU2H(hyp)}  (Fig.~\ref{fig:FSU2HBEST}): This model, which was in good agreement with the observed thermal data without the inclusion of pairing, 
also benefits from pairing. The inclusion of shallow to medium proton pairing, together with neutron superfluidity, leads to an overall 
improved agreement with the observed data. This can be seen in Fig.~\ref{fig:FSU2HBEST} where we plot the results of the FSU2H(hyp) model 
in the case of the medium proton pairing scenario (the red curve in Fig.~\ref{fig:pSC}). 
We note that within this model the best fit to Cas A is obtained for a $1.88 M_{\odot}$ star. 
This is due to the change of slope of the cooling curve by the implementation of the onset of neutron triplet pairing and the PBF process 
within the FSU2H(hyp) model, as seen in the inset of Fig.~\ref{fig:FSU2HBEST}. The results for a $1.4 M_{\odot}$ star are also close to the Cas A 
observations, but the onset of neutron superfluidity, and thus of the PBF processes, happens a little later ($\sim$$10^3$ years),
so that the agreement with the slope of the cooling curve of the Cas A data is not as good. 
In any case, we recall that the Cas A thermal states are still under debate in the literature \citep{Posselt:2013xva}, 
as new stars, contrary to old stars, need carbon atmospheres for their description. Therefore, we have to note that, so far, 
the agreement for Cas A of a certain EoS model and the employed pairing gaps is to be considered with some caution.

One should also notice that at the moment there exist no theoretical calculations of proton pairing in 
neutron-star matter that incorporate consistently all media effects, which subjects the predicted pairing gaps 
to uncertainties in size and in extension. 
We observe in Fig.~\ref{fig:FSU2HBEST} for a $1.88 M_{\odot}$ star that the use of a maximum critical 
temperature for the proton pairing of $1.41 \times 10^9$~K, considerably below the maximum $T_c \sim 10^{10}$~K of our 
assumed parametrizations in Fig.~\ref{fig:pSC}, makes little difference for the cooling behavior of the star.
Most microscopic models of proton singlet pairing, such as those detailed in Fig.~9 of \citep{Page2004},
lead to proton superconductivity extending up to proton Fermi momenta of about 0.8--1.25~fm$ ^{-1}$.
Our phenomenological pairing model is in agreement with several other publications 
\citep{Shternin2011a,2004AstL...30..759G,2005NuPhA.752..590Y,Gusakov2005} that have used
a similar approach. We obtain that the medium proton pairing of the cooling calculations of Fig.~\ref{fig:FSU2HBEST}
extends up to $k_{{\rm F}p}\sim1.3$ fm$ ^{-1}$,  as in the pairing model CCDK---but larger than in the other models---of 
the aforementioned reference \citep{Page2004}. In the shallow pairing scenario (black curve in Fig.~\ref{fig:pSC}), 
we obtain a lower value $k_{{\rm F}p}\sim1.19$~fm$ ^{-1}$ for the maximum extension of proton pairing. The major effect in 
the cooling simulations is that neutrino emission processes are less suppressed, thus making the temperature drop 
sharper and stronger, as can be seen for a $1.88 M_{\odot}$ star in Fig.~\ref{fig:FSU2HBEST}. The modification of 
the cooling curve, however, is not found to be dramatic and does not largely alter our main conclusions.

In summary, while including neutron pairing together with shallow or medium proton pairing is inefficient in slowing the 
thermal evolution of the stars predicted by the FSU2 model that has a stiff nuclear symmetry energy, deep proton pairing does improve
the agreement with data, but still does not explain the cooling of intermediate and low temperature stars.
{The FSU2R(nuc) and FSU2H(hyp) models, which are characterized by a softer symmetry energy than FSU2 and postpone the onset of DU to higher baryon density, already perform well without consideration of pairing. Nevertheless, we know that a complete neglect of pairing in stellar cooling is not realistic, especially for the well-established neutron $^1S_0$ and $^3P_2$ pairing \citep{Page2004}. We find that including medium proton pairing, in addition to the neutron pairing, improves the agreement of the cooling curves of the FSU2R(nuc) and FSU2H(hyp) models with data. This stems from the fact that the proposed pairing scheme suppresses DU processes until about 4 times saturation density. This pairing scheme does not block the nucleonic DU processes in massive stars of the FSU2R(nuc) and FSU2H(hyp) models  (as it takes place at higher densities, see Table \ref{table:cool1}), but suppresses other less efficient cooling mechanisms, such as the modified Urca processes that take place in intermediate-mass stars. In other words, we have found that a shallow/medium proton superconductivity does not lead to an over-suppression of the DU processes, a fact that, combined with the underlying properties of the microscopic model, leads to an optimum agreement with observed data,} especially concerning the slope for the cooling of Cas A in the case of the FSU2H(hyp) model, as can be seen in 
the inset of Fig.~\ref{fig:FSU2HBEST}. This is our preferred model as the colder stars are described with masses 
a little below those for the FSU2R(nuc) model, and also because it allows for the presence of 
hyperons when the chemical equilibrium conditions are fulfilled, being at the same time able of supporting 
neutron stars with 2$M_{\odot}$. 

We note that in our work hyperon pairing has not been taken into account. This was addressed recently
by \cite{2018MNRAS.475.4347R} where the cooling of hyperonic stars was studied, although following a somewhat different
perspective than ours. \cite{2018MNRAS.475.4347R}, employing different relativistic density functional
models, focused their investigation on the role of hyperonic DU processes on cooling, as well as on the effect of pairing on such
processes. They have found that the $\Lambda \rightarrow p + e^- +\bar{\nu}$
is the dominant hyperonic DU process. Furthermore, they have found that
observed data for colder objects can be explained by stars with $M \leq 1.85
M_\odot$ whereas the hotter ones are well explained by stars with $ M \leq
1.6 M_\odot$. Our results are in agreement with their conclusions regarding
the dominant hyperonic DU process. However, in our preferred FSU2H(hyp) model, we have found
that hotter stars can be described by objects with masses of up
to $M \approx 1.85 M_\odot$, as opposed to $ M \leq 1.6 M_\odot$. One must
note that for the aforementioned results \cite{2018MNRAS.475.4347R} have considered pervasive pairing such that
the nucleonic DU process is completely suppressed. The
possibility of nucleonic DU was also considered, but did not produce
agreement with observed data. In this regard, our studies
differ considerably from those of \cite{2018MNRAS.475.4347R}, as we have not completely excluded the nucleonic DU
process and we have not considered hyperonic pairing. For a future
investigation, we wish to expand our study to also include hyperonic pairing.




\section{Conclusions}

We have performed cooling simulations for isolated neutron stars with recently developed equations of state for the core. These are obtained from new parametrizations of the FSU2 relativistic mean-field functional \citep{Chen:2014sca}, which reproduce the properties of nuclear matter and finite nuclei, while fulfilling the restrictions on high-density matter deduced from heavy-ion collisions, measurements of massive 2$M_{\odot}$ neutron stars, and neutron star radii below 13 km.  

The analysis of the cooling behavior has shown that the underlying microscopic model is successful in explaining most of the observed cooling data on isolated neutron stars. Particularly satisfactory is the parametrization FSU2H(hyp), which allows for slow cooling for stars with $M < 1.85 M_\odot$, and for moderate or fast cooling otherwise. This indicates that most of the observed data, as well as Cas A, could be explained, without all observed colder stars being constrained to relatively high mass. A similar behavior is exhibited by the parametrization FSU2R(nuc), however with the caveat that all colder observations would have higher masses above $1.90 M_\odot$ and that hyperons are not present. 

We have also investigated the role of pairing for the cooling behavior of these two models together with the original FSU2 model. 
For that purpose, we have considered phenomenological singlet and triplet neutron pairing, as traditionally assumed for thermal 
evolution calculations,  as well as phenomenological proton pairing in the stellar core. Due to the current uncertainties 
regarding proton pairing,  we have taken into account different ``depths" of proton singlet pairing, i.e.,  we have allowed the protons in the core to pair up to different 
densities in the core, going from low densities (shallow pairing) up to higher densities (deep pairing). As compared to the FSU2 model, we have found that the models FSU2R(nuc) and, 
especially, FSU2H(hyp) only need shallow to moderate proton pairing for a satisfactory agreement with observed cooling data, particularly with Cas A. 
We note, however, that the analysis is subjected to the inherent uncertainties of proton pairing in the core of neutron stars. Therefore, we must be cautious, 
as further developments in microscopic calculations of proton pairing---especially with regards to media polarization effects for proton pairing in the high-density regime---could 
potentially modify such results.

Our calculations indicate that if stellar radii are large (stiff symmetry energy), neutron stars cool down fast for all masses, unless deep proton pairing is active. 
If stellar radii are small (soft symmetry energy), only heavy neutron stars cool down fast, and just shallow to mild proton pairing is needed for improving 
the comparison with the cooling data. The better agreement with the data in the calculations using the FSU2R(nuc) and FSU2H(hyp) models suggests that the cooling observations are more compatible with a soft nuclear symmetry energy and, hence, with small neutron star radii. It is nevertheless to be mentioned that there is a tendency in the present calculations to favor rather large stellar masses for explaining the observed colder  stars with surface temperatures $T \lesssim 10^6$ K. 

As for future perspectives, apart from the inclusion of hyperonic pairing, we intend to extend our study of the thermal evolution of neutron stars within this microscopic model to the analysis
of the influence of rotation and magnetic fields. Both these effects are known to break the spherical symmetry of the star and could influence the microscopic, macroscopic and thermal properties of the star \citep{Negreiros2012,Negreiros2013,Negreiros2017}.

\section*{Acknowledgements}
R.N. acknowledges financial support from CAPES and CNPq, as well as  that this work is a 
part of the project INCT-FNA Proc. No. 464898/2014-5.
L.T. acknowledges support from the Ram\'on y Cajal research programme,
FPA2013-43425-P and FPA2016-81114-P Grants from Ministerio de Econom\'{\i}a y 
Competitividad (MINECO), Heisenberg Programme of the Deutsche Forschungsgemeinschaft under the Project Nr. 383452331 and PHAROS COST Action CA16214.
M.C. and A.R. acknowledge support from Grants No. FIS2014-54672-P and No. 
FIS2017-87534-P from MINECO, and the project MDM-2014-0369 of ICCUB (Unidad de 
Excelencia Mar\'{\i}a de Maeztu) from MINECO.

\bibliography{biblio}

\end{document}